\documentclass[twocolumn]{aastex631}

\usepackage{amssymb}
\usepackage{amsmath}

\received{October, 2022}
\revised{April, 2023}
\accepted{May, 2023}

\submitjournal{The Astrophysical Journal}

\shorttitle{AASTeX v6.3.1 Sample article}
\shortauthors{Steinrueck et al.}

\begin{document}

\title{Photochemical hazes dramatically alter temperature structure and atmospheric circulation in 3D simulations of hot Jupiters}

\correspondingauthor{Maria E. Steinrueck}
\email{steinrueck@mpia.de}

\author[0000-0001-8342-1895]{Maria E. Steinrueck}
\affiliation{Max-Planck-Institut f\"ur Astronomie, 69117 Heidelberg, Germany}
\affiliation{Lunar and Planetary Laboratory, University of Arizona, Tucson, AZ, 85721, USA}

\author[0000-0003-3071-8358]{Tommi Koskinen}
\affiliation{Lunar and Planetary Laboratory, University of Arizona, Tucson, AZ, 85721, USA}

\author[0000-0002-5360-3660]{Panayotis Lavvas}
\affiliation{Groupe de Spectrom\'etrie Moleculaire et Atmosph\'erique, Universit\'e de Reims Champagne-Ardenne, Reims, France}

\author[0000-0001-9521-6258]{Vivien Parmentier}
\affiliation{Atmospheric, Ocean, and Planetary Physics, Department of Physics, Oxford University, OX1 3PU, UK}
\affiliation{Universit\'e C\^ote d’Azur, Observatoire de la C\^ote d’Azur, CNRS, Laboratoire Lagrange, France}

\author[0000-0003-0562-6750]{Sebastian Zieba}
\affiliation{Max-Planck-Institut f\"ur Astronomie, 69117 Heidelberg, Germany}

\author[0000-0003-2278-6932]{Xianyu Tan}
\affiliation{Atmospheric, Ocean, and Planetary Physics, Department of Physics, Oxford University, OX1 3PU, UK}

\author[0000-0002-8706-6963]{Xi Zhang}
\affiliation{Department of Earth and Planetary Sciences, University of California, Santa Cruz, CA 95064, USA}

\author[0000-0003-0514-1147]{Laura Kreidberg}
\affiliation{Max-Planck-Institut f\"ur Astronomie, 69117 Heidelberg, Germany}



\begin{abstract}
Photochemical hazes are expected to form in hot Jupiter atmospheres and may explain the strong scattering slopes and muted spectral features observed in the transmission spectra of many hot Jupiters. Absorption and scattering by photochemical hazes have the potential to drastically alter temperature structure and atmospheric circulation of these planets but have previously been neglected in general circulation models (GCMs). We present GCM simulations of hot Jupiter HD 189733b that include photochemical hazes as a radiatively active tracer fully coupled to atmospheric dynamics. The influence of haze radiative feedback strongly depends on the assumed haze optical properties. For soot hazes, two distinct thermal inversions form, separated by a local temperature minimum around 10$^{-5}$ bar caused by upwelling on the dayside mixing air with low haze abundance upwards. The equatorial jet broadens and slows down. The horizontal distribution of hazes remains relatively similar to simulations with radiatively passive tracers. For Titan-type hazes, the equatorial jet accelerates and extends to much lower pressures, resulting in a dramatically different 3D distribution of hazes compared to radiatively passive or soot hazes. Further experimental and observational studies to constrain the optical properties of photochemical hazes will therefore be crucial for understanding the role of hazes in exoplanet atmospheres. In the dayside emission spectrum, for both types of hazes the amplitude of near-infrared features is reduced, while the emitted flux at longer wavelengths ($>$4 $\mu$m) increases. Haze radiative feedback leads to increased phase curve amplitudes in many infrared wavelength regions, mostly due to stronger dayside emission.
\end{abstract}



\section{Introduction} \label{sec:intro}
Transit observations of many short-period giant planets reveal the presence of aerosols at low pressures \citep{SingEtAl2016,CrossfieldKreidberg2017,GaoEtAl2021Review}. Among the observed spectral signatures of aerosols are short-wavelength scattering slopes \citep[e.g.,][]{PontEtAl2008,PontEtAl2013,NikolovEtAl2015WASP-6b,AlamEtAl2020HAT-P-32b}, muted wings of the sodium and potassium lines \citep[e.g.,][]{HuitsonEtAl2012,GibsonEtAl2013HAT-P-32b,MallonnStrassmeier2016}  and the low amplitude of the near-infrared water feature near 1.4 $\mu$m \citep[e.g.,][]{LineEtAl2013HAT-P-12b,DemingEtAl2013HD209458b_XO-1b,McCulloughEtAl2014,IyerEtAl2016H2Ofeature,WakefordEtAl2017WASP-101b}. In some cases, there is evidence for an aerosol layer spanning many pressure scale heights \citep{PontEtAl2013,EstrelaEtAl2021WASP-69b}, requiring aerosols to be present at pressures as low as 1 $\mu$bar \citep{EstrelaEtAl2021WASP-69b}.

Two fundamentally different formation mechanisms for these aerosols have been proposed: particles forming through condensation of gases as they are transported towards cooler regions of the atmosphere (condensate clouds) and particles forming through a complex chain of photochemical reactions initiated by UV light at high altitudes (photochemical hazes). While condensate clouds are the most likely type of aerosol in most of the hotter hot Jupiters \citep[e.g.,][]{SudarskyEtAl2000,WakefordEtAl2015,PowellEtAl2018,GaoEtAl2020}, photochemical hazes are thought to dominate over condensate clouds for cooler planets, especially at high altitudes. The exact temperature of the transition is model-dependent. While \citet{GaoEtAl2020} find photochemical hazes to be the dominant source of opacity for equilibrium temperatures $< 950$ K, \citet{LavvasKoskinen2017} predict that photochemical hazes could explain the transmission spectrum of HD 189733b ($T_{\mathrm{eq}} \approx 1,200$ K). \citet{ArfauxLavvas2022} found that the upper temperature limit based on the observations is between 1400 and 1700~K. In laboratory experiments, \citet{FleuryEtAl2019} observed that photochemical hazes could form in hydrogen-dominated atmospheres as hot as 1500 K if the C/O ratio is supersolar.

In addition, it has been proposed that photochemical hazes could explain “super-Rayleigh slopes” (scattering slopes that are steeper than what would be expected for Rayleigh scattering with a constant abundance of scatterers) more naturally than condensate clouds \citep{PontEtAl2013,OhnoKawashima2020}. Thus, photochemical hazes can be important for explaining the optical and UV spectrum even for planets in which condensate clouds dominate the infrared opacity.

It has been well-established that aerosols have a strong potential to alter the atmospheric temperature and circulation, as known from examples in the Solar System: On Titan, absorption and scattering by hazes are important contributions to the energy budget of the atmosphere. Hazes create a thermal inversion at low pressures and have a cooling effect on deeper atmospheric regions and the surface, called the anti-greenhouse effect \citep{McKayEtAl1991}. Coupling a haze microphysics model with a general circulation model has been crucial for explaining the observed haze structure and circulation of Titan \citep{RannouEtAl2002,LebonnoisEtAl2012}. 

For extrasolar giant planets, multiple studies on the radiative effects of condensate clouds in GCMs of hot Jupiters establish that radiative feedback from aerosols is significant. Here, we briefly review these papers, sorted roughly from the least complex model to the most complex one. \citet{OreshenkoEtAl2016} examined the effect of including scattering in a double-gray model of a hot Jupiter assuming uniform scattering properties throughout the atmosphere. \citet{RomanRauscher2017PrescribedClouds} took a similar approach but increased the complexity by prescribing different static, horizontally inhomogeneous cloud coverages motivated by optical phase curve observations of Kepler-7b. For their vertical cloud coverage, they assumed that the cloud would extend from a chosen cloud base to the top of the model, with a constant mixing ratio. They found that inhomogeneous clouds significantly impacted the temperature structure as well as the equatorial jet, but that the prescribed static cloud coverages resulted in a simulation that was not energy-balanced. In a follow-up study, \citet{RomanRauscher2019} updated their model to include a physically motivated cloud location, such that clouds form in any atmospheric column in which the temperature profile crosses the condensation curve of a relevant cloud species. Based on these models, \citet{HaradaEtAl2021} found that radiative feedback from clouds significantly affected high-resolution spectra of hot Jupiters. \citet{ParmentierEtAl2016} used an approach similar to \citet{RomanRauscher2019} but using wavelength-dependent radiative transfer, though the results of their simulations including haze feedback are only briefly discussed in their publication. In \citet{RomanEtAl2021} and \citet{ParmentierEtAl2021}, their respective models were applied to a much larger range of equilibrium temperatures.
\citet{LinesEtAl2019} and \citet{ChristieEtAl2021} also employed a similar though somewhat more complex approach, calculating cloud properties such as the vertical distribution and particle size based on 1D cloud model EDDYSED \citep{AckermanMarley2001}.

A different and dynamically more self-consistent approach is to include one or several cloud species as a tracer in the model, thus simulating how clouds are transported within the atmosphere. After first studies modeling clouds as passive tracers \citep{ParmentierEtAl2013,CharnayEtAl2015a}, neglecting radiative feedback, \citet{CharnayEtAl2015b} were the first to model radiatively active tracers representing clouds on a short-period extrasolar giant planet (in their case, a mini-Neptune). In their model, all material exceeding the vapor pressure condensed into particles with a prescribed, fixed size. Heating by clouds produced a dayside thermal inversion, the strength of which was limited by the evaporation of the cloud \citep[an effect also observed in the local equilibrium cloud model of][]{RomanRauscher2019}. Cloud radiative feedback in addition led to a more severe depletion of clouds in the equatorial zone compared to higher latitudes. More recently, \citet{KomacekEtAl2022UHJclouds} simulated radiatively active tracer clouds using a comparable model in the atmospheres of ultra-hot Jupiters, showing that cloud patchiness might lead to a higher thermal emission on the nightside. Finally, both \citet{LeeEtAl2016} and \citet{LinesEtAl2018} coupled a full microphysics model to a GCM of a hot Jupiter. Their model traces the abundances of multiple gas species and captures the key processes of nucleation, particle growth of mixed-species grains through surface reactions, and evaporation in addition to transport and gravitational settling of cloud particles. It is by far the most complex cloud model that has been applied to extrasolar giant planets. In both studies, heating and cooling by clouds had a significant effect on temperature structure, cloud abundance and atmospheric circulation. Comparing the results between both studies, \citet{LinesEtAl2018} further concluded that explicit treatment of scattering \citep[as opposed to adding the scattering cross section to the absorption cross section, as done in ][]{LeeEtAl2016} is important.

Given the established significance of radiative feedback for condensate clouds on short-period giant planets, the role of photochemical hazes on these atmospheres is less well-studied. Studies using one-dimensional radiative transfer models have found that absorption and scattering by photochemical hazes can lead to stark changes in the temperature profile, similar to the anti-greenhouse effect on Titan. For mini-Neptunes, \citet{MorleyEtAl2015} found that soot-based photochemical hazes created a thermal inversion of up to 200 K at low pressures, while simultaneously cooling deeper layers of the atmosphere by several hundred Kelvin. This temperature inversion led to emission spectra that substantially differed from models of a clear atmosphere or an atmosphere with condensate clouds. More recently, \citet{LavvasArfaux2021} and \citet{ArfauxLavvas2022} examined haze radiative feedback in hot Jupiter atmospheres. They confirmed that in this case, a thermal inversion also formed at low pressures, with the detailed temperature structure depending on the refractive index of the hazes. However, the feedback of photochemical hazes on the atmospheric circulation has not yet been studied with general circulation models.

The goal of this work is to investigate the role of radiative feedback of photochemical hazes in the atmospheres of hot Jupiter exoplanets. Mass mixing ratios of photochemical hazes generally peak at lower pressures than condensate clouds, suggesting that hazes could affect atmospheric dynamics differently from clouds.
First simulations of the 3D distribution of photochemical hazes, modeling hazes as a radiatively passive tracer, demonstrate that a complex and highly inhomogeneous global distribution can be expected \citep{SteinrueckEtAl2021}.  Motivated by these findings, we add complexity to the haze model of \citet{SteinrueckEtAl2021} by coupling the haze model to the radiative transfer, thus adding heating and cooling by hazes to the dynamics. In addition, we switch to using wavelength-dependent radiative transfer \citep{ShowmanEtAl2009,KatariaEtAl2013} rather than the previously used double-gray radiative transfer to model more realistic heating and cooling rates. In terms of the level of complexity and modeling approach, our model can thus be viewed as a photochemical haze equivalent to the \citet{CharnayEtAl2015b} model. As in \citet{SteinrueckEtAl2021}, we focus on HD~189733b, which is one of the best-characterized exoplanets to date and which shows evidence of aerosols in its transmission spectrum \citep[e.g.,][]{TinettiEtAl2007HD189WaterVapor,KnutsonEtAl2007HD189phasecurve,PontEtAl2008,GibsonEtAl2012HD189,KnutsonEtAl2012,MajeauEtAl2012, McCulloughEtAl2014,LoudenWheatley2015,AngerhausenEtAl2015SOFIA,BrogiEtAl2016,FlowersEtAl2019,SeidelEtAl2020HD189,SanchezLopezEtAl2020,KingEtAl2021}.

While this paper focuses on hot Jupiters, for which higher quality observations are available, we anticipate that our work will also lay the groundwork for later studies of haze radiative feedback on cooler and smaller tidally locked giant planets, for which photochemical hazes are predicted to form efficiently \citep{MorleyEtAl2015,HorstEtAl2018,HeEtAl2018,KawashimaIkoma2019,AdamsEtAl2019,LavvasEtAl2019} and for which there is ample observational evidence of aerosols in transmission spectra \citep{CrossfieldKreidberg2017}.

The remainder of the paper is structured as follows: Section \ref{sec:methods} describes our model. In Section \ref{sec:grayvssparc}, we compare simulation results from the double-gray model used in \citet{SteinrueckEtAl2021} to results using a wavelength-dependent model using the correlated-k method. Section \ref{sec:hazefeedback} describes simulations with haze radiative feedback, with Section \ref{subsec:soothazes} focusing on a refractive index of soot and Section \ref{subsec:titanhazes} focusing on a refractive index similar to Titan-type hazes. In Section \ref{sec:observations}, we explore the impact on model-predicted transmission and emission spectra as well as phase curves. Finally, we discuss caveats and directions for future work in Section \ref{sec:discussion} and summarize our findings in Section \ref{sec:conclusion}.

\section{Methods}\label{sec:methods}
We use SPARC/MITgcm to simulate the atmosphere of hot Jupiter HD~189733b. SPARC/MITgcm couples the plane-parallel, wavelength-dependent radiative transfer code of \citet{MarleyMcKay1999} to the general circulation model of \citet{AdcroftEtAl2004}. 
It has been applied to a wide range of hot Jupiters and other exoplanets \citep[e.g.,][]{ShowmanEtAl2009,ShowmanEtAl2013DopplerSignatures,LewisEtAl2014,KatariaEtAl2015,KatariaEtAl2016,SteinrueckEtAl2019,ParmentierEtAl2018,ParmentierEtAl2021}. In addition, to facilitate comparison with \citet{SteinrueckEtAl2021}, we also include one simulation using MITgcm with double-gray radiative transfer.

\subsection{Atmospheric Dynamics}

\begin{deluxetable}{lrr}




\tablecaption{Model Parameters}
\label{tab:modelparameters}
\tablehead{\colhead{Parameter} & \colhead{Value} & \colhead{Units} \\ 
\colhead{} & \colhead{} & \colhead{} } 

\startdata
		Radius\tablenotemark{1,2}& $1.13$ & $R_{\textrm{Jup}}$ \\ 
		Gravity\tablenotemark{1}& 21.93 & m s$^{-2}$ \\ 
		Rotation period\tablenotemark{1} & 2.21857567  & d \\
		Semimajor axis\tablenotemark{3} & 0.03142 & AU \\
		Specific heat capacity & $1.3\cdot 10^4$ & J kg$^{-1}$ K$^{-1}$ \\
		Specific gas constant & 3714 & J kg$^{-1}$ K$^{-1}$ \\
		Horizontal resolution & C32\tablenotemark{a} &  \\
		Vertical resolution & 60 & layers \\
		Lower pressure boundary & $1.75 \cdot 10^{-7}$ & bar \\
		Upper pressure boundary & 200 & bar \\
		Temperature of bottom-most layer\tablenotemark{4} & 2891 & K \\ 
		Hydrodynamic time step & 25 & s \\
		Radiative time step \tablenotemark{5} & 50 & s \\
\enddata

\tablenotetext{1}{\citet{StassunEtAl2017}}
\tablenotetext{2}{$R_{\textrm{Jup}}$ here denotes the nominal equatorial radius of Jupiter, with a value of  $7.1492 \cdot 10^7$~m, as defined by IAU 2015 Resolution B3.}
\tablenotetext{3}{\citet{SouthworthEtAl2010III}}
\tablenotetext{4}{The center of the bottom-most layer is located at 170 bar. Temperatures are defined at the center of layers.}
\tablenotetext{5}{In the soot simulation with a haze production rate of $2.5\cdot 10^{-11}$~kg~m$^{-2}$~$s^{-1}$, a radiative time step of 25~s was used. }
\tablenotetext{a}{equivalent to a resolution of 128x64 on a longitude-latitude grid}


\end{deluxetable}

We solve the global primitive equations on a cubed-sphere grid using the MITgcm in its atmosphere configuration. The primitive equations are an approximation of the fluid dynamics equations 
that is valid for stably-stratified shallow atmospheres. It has been demonstrated that they are a good approximation when simulating the atmospheres of hot Jupiters \citep{ShowmanGuillot2002,MayneEtAl2014}.
The simulation parameters are summarized in Table \ref{tab:modelparameters}. 

We use a fourth-order Shapiro filter to suppress small numerical fluctuations at the grid scale that otherwise could grow and cause instabilities.
Similar to \citet{LiuShowman2013}, we include a drag in the deep atmosphere. This both stabilizes the simulation and ensures independence of the initial condition. The form of the drag is given by $k_v=k_F (p - p_{\textrm{drag,top}})/(p_{\textrm{bottom}}-p_{\textrm{drag,top}})$, where $p_{\textrm{bottom}}$ is the bottom boundary of the simulation domain (200 bar). \citet{CaroneEtAl2020WASP-43b} found that a bottom boundary of 200 bar is sufficiently deep for planets with a rotation period $\gtrapprox$ 1.5 days, well-fulfilled by HD~189733b. We choose $k_F=10^{-4}$ s$^{-1}$ and $p_{\textrm{drag,top}}=10$ bar.

\citet{ThorngrenEtAl2019InternalTemperature} suggested that based on the observed distribution of hot Jupiter radii, the internal heat flux in most hot Jupiters likely is significantly higher than frequently assumed in GCMs. As a consequence, the radiative-convective boundary also is shallower, reaching into typical simulation domains of GCMs. We therefore changed the treatment of the bottom boundary condition compared to our previous model \citep{SteinrueckEtAl2021}, where a uniform net flux was prescribed at the bottom. Instead, we assume that the deepest model layers have reached the convective zone and relax the temperature at the bottom-most layer towards a prescribed value, with a relaxation timescale of $10^5$ s. This treatment, similar to \citet{MayKomacekEtAl2021}, crudely mimics the effect of convective mixing in controlling the deep temperature structure in our model. The temperature of the bottom-most layer at 170 bars, 2891~K, was chosen by interpolating temperature profiles from the grid of models presented in \citet{ThorngrenEtAl2019InternalTemperature} to the gravity and incident flux of HD~189733b. The intrinsic temperature corresponding to this temperature profile is $\approx375$~K. Further, we include a convective adjustment scheme based on the dry adiabatic adjustment scheme used in the Community Atmosphere Model \citep[CAM,][p. 100]{CollinsEtAl2004CAMDocumentation} in our simulations. We found that in addition to being physically motivated, these changes lead to improved numerical stability at long simulation runtimes.

\subsection{Radiative Transfer}
\subsubsection{Wavelength-dependent radiative transfer}
\label{subsec:methods_SPARC}
The radiative transfer used in SPARC/MITgcm is based on the plane-parallel, two-stream radiative transfer code by \citet{MarleyMcKay1999}, that was originally developed for Titan \citep{McKayEtAl1989} and later adapted for brown dwarfs and exoplanets \citep[e.g.,][]{MarleyEtAl1996,FortneyEtAl2005,FortneyEtAl2008,MorleyEtAl2012}. It was first coupled to the MITgcm by \citet{ShowmanEtAl2009}. 
We use the version with 11 wavelength bins introduced by \citet{KatariaEtAl2013}, which is optimized for computational speed while maintaining accuracy. The code uses the correlated-k method \citep[e.g.,][]{GoodyYung1989} to describe molecular opacities within each wavelength bin. Correlated-k coefficients are calculated assuming abundances based on the equilibrium chemistry calculations of \cite{LoddersFegley2002} and \cite{VisscherEtAl2006}, assuming solar elemental abundances. Molecular opacities are taken from \citet{FreedmanEtAl2008}, including the updates from \citet{FreedmanEtAl2014}. 
We note that our previous work \citep{SteinrueckEtAl2019} as well as \citet{DrummondEtAl2018HD189733b} and \citet{DrummondEtAl2020} found that on HD~189733b, transport-induced disequilibrium abundances of CH$_4$ and H$_2$O can alter temperatures in the lower atmosphere by up to 10\% compared to equilibrium chemistry. However, the method of \citet{SteinrueckEtAl2019}, who assumed homogeneous CH$_4$, CO and H$_2$O abundances throughout the atmosphere,  is only a valid approximation for pressures above $\approx 10^{-4}$ bar. At lower pressures, photochemistry destroys CH$_4$, leading to a rapid drop in the CH$_4$ abundances with decreasing pressure \citep{MosesEtAl2011}. Assuming a constant CH$_4$ abundances as in \citet{SteinrueckEtAl2019} thus considerably overestimates CH$_4$ abundances at low pressures. Incidentally, at these low pressures, equilibrium chemistry predicts low CH$_4$ abundances that rapidly decline with decreasing pressure, closer to what is observed in 1D models that include photochemistry \citep{MosesEtAl2011}, despite underpredicting CH$_4$ abundance at pressures between $\approx10^{-4}-1$ bar.
Because this paper focuses on the temperature, circulation and haze distribution at these low pressures, we assume equilibrium abundances for all species for simplicity. We note that this assumption does not affect the haze production rate in our model, which we treat as a free parameter.

\begin{figure}
\begin{center}
\includegraphics[width=0.95\columnwidth]{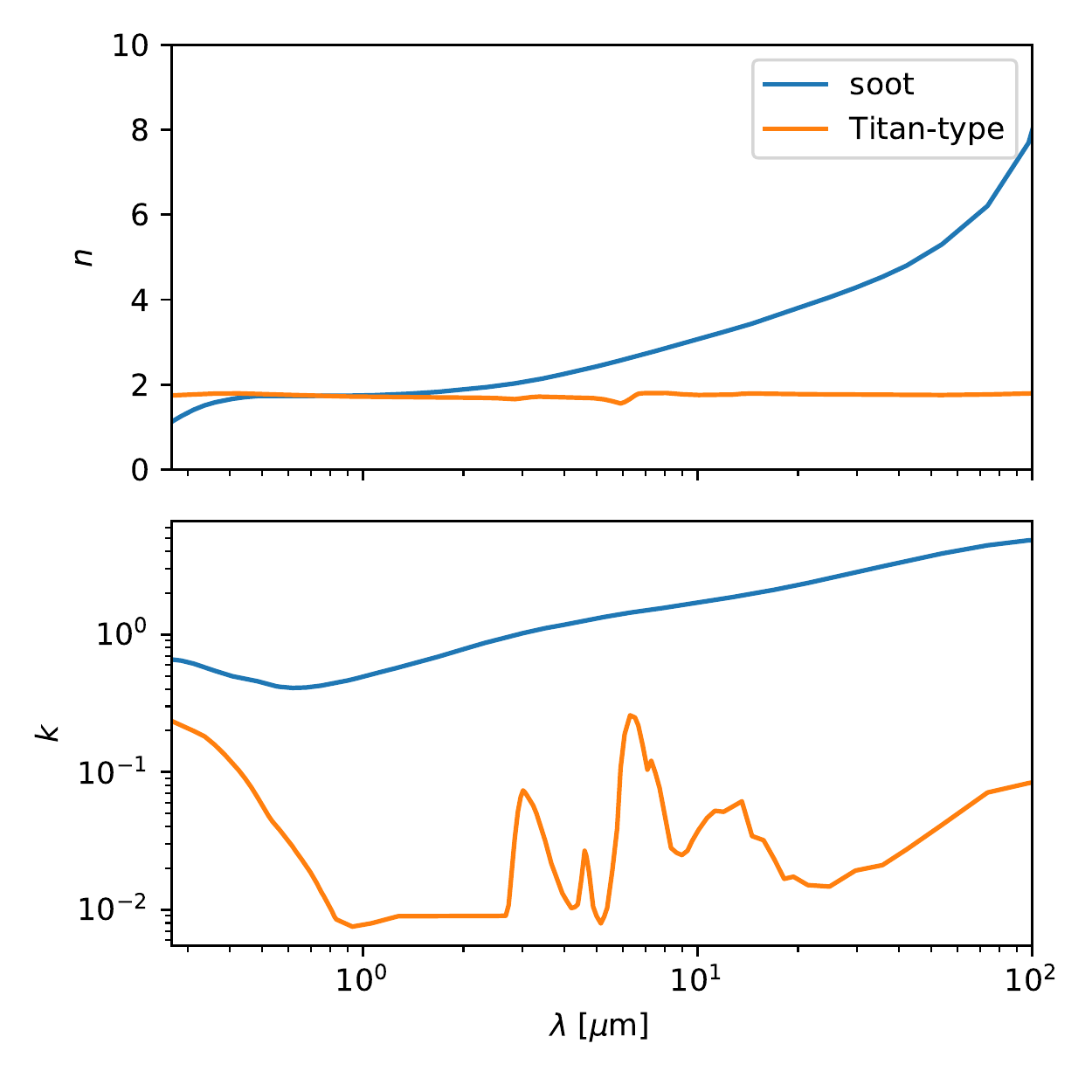}
\caption{Real (top panel) and imaginary (bottom panel) parts of the complex refractive indices used in this work, shown at the 196-wavelength-bin resolution used for post-processing.}
\label{fig:refractiveindex}
\end{center}
\end{figure}

Absorption and scattering by hazes is calculated based on Mie theory \citep{Mie1908}. In order to smooth the Mie oscillations that would be observed for a single particle size, we use a narrow log-normal with a geometric standard deviation of 1.05 for the particle size distribution within the radiative transfer code \citep[similar to ][]{ParmentierEtAl2016,ParmentierEtAl2021}. The haze opacity is linearly related to the local haze abundance. The haze abundance used in the radiative transfer calculation is directly coupled to the time- and location-dependent tracer describing the haze mass mixing ratio (see below). 
The refractive index of the haze particles, an important input quantity for our calculations, is poorly constrained. In the absence of laboratory measurements specifically conducted with exoplanets in mind, soots frequently have been used as analog for high-temperature hazes \citep[e.g.,][]{MorleyEtAl2013,MorleyEtAl2015,LavvasKoskinen2017, OhnoKawashima2020,LavvasArfaux2021,SteinrueckEtAl2021}. We therefore assume refractive indices of haze particles based on measurements of soot particles formed in combustion experiments for our nominal simulations. Specifically, we use the refractive indices from \citet{LavvasKoskinen2017}, who combine the measurements from several different groups \citep{LeeTien1981,ChangCharalampopoulos1990,GavilanEtAl2016} in order to cover a broad wavelength range. While the detailed complex refractive index can vary between different soot experiments \citep[e.g.,][]{JaegerEtAl1999SpectralPropertiesOfCarbonBlack}, soots are in general known to be highly absorbing over a broad wavelength range. To explore the effect of a different composition of the hazes, we also ran simulations using refractive indices typical of Titan hazes. Here, we use the refractive index from \citet{LavvasEtAl2010}, who base the real part of the refractive index on laboratory experiments simulating haze formation on Titan \citep{KhareEtAl1984} and retrieve the imaginary part from observations with the \textit{Descent Imager/Spectral Radiometer} (DISR) of the \textit{Huygens} probe. 
We note that both Titan's haze (collected by the \textit{Huygens} probe) as well as laboratory Titan haze analogs (tholins) pyrolyze at temperatures above $\approx600$~K \citep{IsraelEtAl2005,MorissonEtAl2016} and thus are an unlikely candidate for hazes in hot Jupiter atmospheres. However, given the lack of knowledge of the optical properties of hazes in hot Jupiter atmospheres, it is useful to consider Titan-type hazes as an example of hazes that are more reflective and have a stronger wavelength-dependence of the extinction cross-sections than soots. Titan-type haze refractive indices have been used in this sense in multiple other studies of hot Jupiters \citep{OhnoKawashima2020,LavvasArfaux2021}. We discuss the limitations of the choice of the refractive indices in more detail in Section \ref{sec:discussion}.
 The refractive indices for soots and Titan-type hazes used in this work are identical to the ones used by \citet{LavvasArfaux2021} and are shown in Fig. \ref{fig:refractiveindex}.

\subsubsection{Double-gray radiative transfer}
In addition to the simulations using SPARC, we also include one simulation using the double-gray radiative transfer used in \citet{SteinrueckEtAl2021} for comparison. In this simulation, the TWOSTR package \citep{KyllingEtAl1995}, which is based on the multistream discrete ordinate algorithm DISORT \citep{StamnesEtAl1988},  is used to solve the radiative transfer equations for a plane-parallel atmosphere in the two-stream approximation. For the opacities, we choose a value of $\kappa_{v}=6 \cdot 10^{-4} \sqrt{T_{\mathrm{irr}}/2000 \text{K}}$ m$^2$ kg$^{-1} =5.5 \cdot 10^{-4}$ m$^2$ kg$^{-1}$ in the visible band and a value of $\kappa_{\mathrm{th}}=10^{-3}$ m$^2$ kg$^{-1}$ in the thermal band  \citep{ParmentierGuillot2014}. Except for the choice of the bottom boundary condition and initial condition, which are chosen to be identical to the other simulations in this work, the simulation setup is identical to \citet{SteinrueckEtAl2021}. 

\subsection{Haze model}
Photochemical hazes are included in the GCM as a tracer. 
The haze mass mixing ratio $\chi$ obeys
\begin{equation}
\frac{D \chi}{Dt} = -g \frac{\partial (\rho \chi V_s)}{\partial p} + P + L,
\label{eq:tracereqn}
\end{equation}
where $D/Dt$ is the material derivative $\partial/\partial t + \bf{v}_H \cdot \nabla_H + \omega \partial/\partial p$, with $\bf{v}_H$ being the horizontal velocity, $\bf{\nabla}_H$ the horizontal gradient operator on a sphere in pressure coordinates and $\omega$ the vertical velocity in pressure coordinates. Furthermore, $g$ is the gravitational acceleration, $\rho$ is the gas density and $V_s$ is the settling velocity of the haze particles in the atmosphere in m~s$^{-1}$. 
For the production term $P$, we assume a log-normal distribution in pressure,
\begin{equation}
    P= F_0 \, g \cos \theta \cdot \frac{1}{\sqrt{2 \pi} p \sigma} \: \exp \left( - \frac{\ln^2(p/m)}{2 \sigma^2}\right),
\end{equation}
with a median $m=2$ $\mu$bar and a standard deviation $\sigma=0.25 \ln(10)\approx 0.576$. Here, $F_0$ is the column-integrated haze mass production rate at the substellar point (given in Table \ref{tab:simulationsoverview}) and $\theta$ is the angle of incidence of the starlight. The parameters of the distribution were chosen such that haze production is negligible in the two top-most layers. We note that except for the value of $F_0$, this production term is identical to the production term used in \citet{SteinrueckEtAl2021}, though we here choose to write it directly as a function of $p$ for improved clarity.

The loss term $L$ is given by 
\begin{equation}
L=  
\begin{cases}
	0 & \text{for } p<p_{\mathrm{deep}}, \\
	-\chi/\tau_{loss} & \text{for } p>p_{\mathrm{deep}} ,
\end{cases}
\end{equation}
with the loss time scale $\tau_{\mathrm{loss}}=10^{3}$ s and $p_{\mathrm{deep}}=100$~mbar. This term is an idealized representation of the condensation of cloud species on top of the haze particles, thus removing them from the distribution of pure hazes, as well as the thermal destruction of hazes in the deep atmosphere. The particular value of $p_{\mathrm{deep}}$ was chosen for numerical reasons: In tests exploring the sensitivity of the 3D haze distribution to the choice of $p_{\mathrm{deep}}$, larger values led to longer convergence times for the haze distribution. However, the final haze distribution for $p\ll p_{\mathrm{deep}}$ did not substantially depend on $p_{\mathrm{deep}}$. A more detailed description of the model can be found in \citet{SteinrueckEtAl2021}. For the simulations presented here, we fix the particle size to 3~nm, close to particle size in 1D~microphysics models of photochemical hazes in the atmosphere of HD~189733b \citep{LavvasKoskinen2017}, and assume a particle density of 1,000~kg~m$^{-3}$.

\subsection{Simulation runtime and overview of the simulations}

\begin{deluxetable*}{llrl}




\tablecaption{List of simulations}
\label{tab:simulationsoverview}

\tablehead{\colhead{Radiative transfer} & \colhead{Haze feedback} & \colhead{Haze production rate\tablenotemark{1}} & \colhead{Refractive index} \\ 
\colhead{} & \colhead{} & \colhead{(kg~m$^{-2}$~$s^{-1}$)} & \colhead{} } 

\startdata
double-gray & off & $2.5\cdot 10^{-12}$ & N/A \\
correlated-k & off& $2.5\cdot 10^{-12}$ & N/A \\
correlated-k & on & $2.5\cdot 10^{-12}$ & soot \\
correlated-k & on & $5\cdot 10^{-12}$ & soot \\
correlated-k & on & $1\cdot 10^{-11}$ & soot \\
correlated-k & on & $2.5\cdot 10^{-11}$ & soot \\
correlated-k & on & $2.5\cdot 10^{-11}$ & Titan-type \\
correlated-k & on & $1\cdot 10^{-10}$ & Titan-type \\
\enddata

\tablenotetext{1}{ at substellar point, column-integrated}


\end{deluxetable*}

Table \ref{tab:simulationsoverview} provides an overview of the simulations. All simulations were initiated from a state of rest with an initial temperature profile interpolated from the grid of \citet{ThorngrenEtAl2019InternalTemperature} and run for 4,500 Earth days simulation time. The simulation time necessary for convergence of the haze distribution depends on two factors: How fast hazes are transported downward and how long it takes to produce the amount of hazes present in the equilibrium state. The former is determined by the smaller one of the vertical mixing timescale and the gravitational settling timescale. For small particle sizes, the vertical mixing timescale is shorter except at very low pressures. The vertical mixing time scale $\tau_{\mathrm mix} = H^2/K_{\mathrm zz}$ \citep[estimated using Eq. 9 in][]{SteinrueckEtAl2021} varies from less than an hour at 1~$\mu$bar to $\approx900$~days at 100~mbar, and thus is not the limiting factor for convergence. The simulation runtime was therefore chosen by monitoring the total mass of hazes in the simulation until it reached a quasi-steady state. We note that in the quasi-steady state, the total mass of hazes still fluctuated by up to 10\% over timescales of a few hundred days.
Unless stated otherwise, our simulation results stated below have been averaged over the last 100 days of simulation time.

\subsubsection{Transit spectra}
To obtain transit spectra, we use a one-dimensional line-by-line radiative transfer code.  To account for inhomogeneities at the terminator, we calculate the transmission spectrum separately for the morning and evening  terminator as well as for the combined effect of the two limbs. Molecular and atomic species included are H$_2$O, CH$_4$, CO, CO$_2$, Na and K. The code further includes Rayleigh scattering by H$_2$ and collision-induced absorption by H$_2$-H$_2$ and H$_2$-He pairs. We treat the haze particles using Mie scattering with the same complex refractive indices as in the GCM. We choose the reference pressure such that the planet radius in the Spitzer 3.6~$\mu$m band matches the observations at that wavelength. Detailed descriptions of the code and opacities used can be found in \citet{LavvasKoskinen2017} and \citet{LavvasArfaux2021}.
\subsubsection{Reflection spectra, emission spectra, and phase curves}
We calculate reflection spectra, emission spectra and phase curves using the same radiative transfer code and opacity sources as in the GCM with wavelength-dependent radiative transfer (Section \ref{subsec:methods_SPARC}), however, using 196 frequency bins. At each orbital phase,  the radiative transfer equation is solved along the line of sight for each atmospheric column. The outgoing fluxes then are combined by performing a weighted average across the disk that is visible from Earth at the given phase.
For the star, we use a NextGen spectrum \citep{HauschildtEtAl1999} and a stellar radius of 0.805 $R_{\odot}$ \citep{BoyajianEtAl2015}.

\section{Passive Tracer Simulations: Double-gray vs correlated-k\label{sec:grayvssparc}}
\begin{figure}
\begin{center}
\includegraphics[width=\columnwidth]{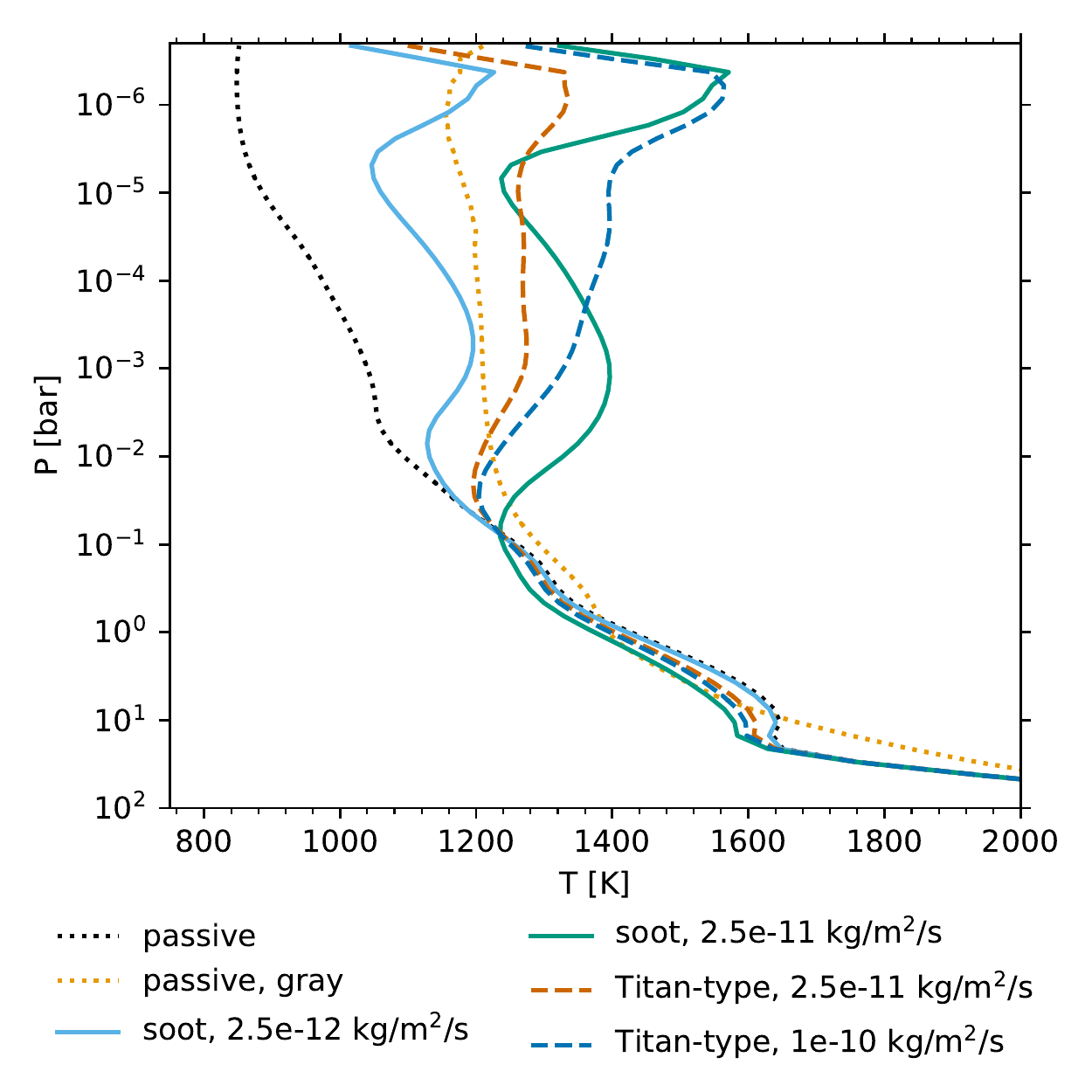}
\caption{Dayside temperature profiles, calculated using an average weighted by the cosine of the angle of incidence.}
\label{fig:temperatureaverages_day}
\end{center}
\end{figure}

\begin{figure}
\begin{center}
\includegraphics[width=\columnwidth]{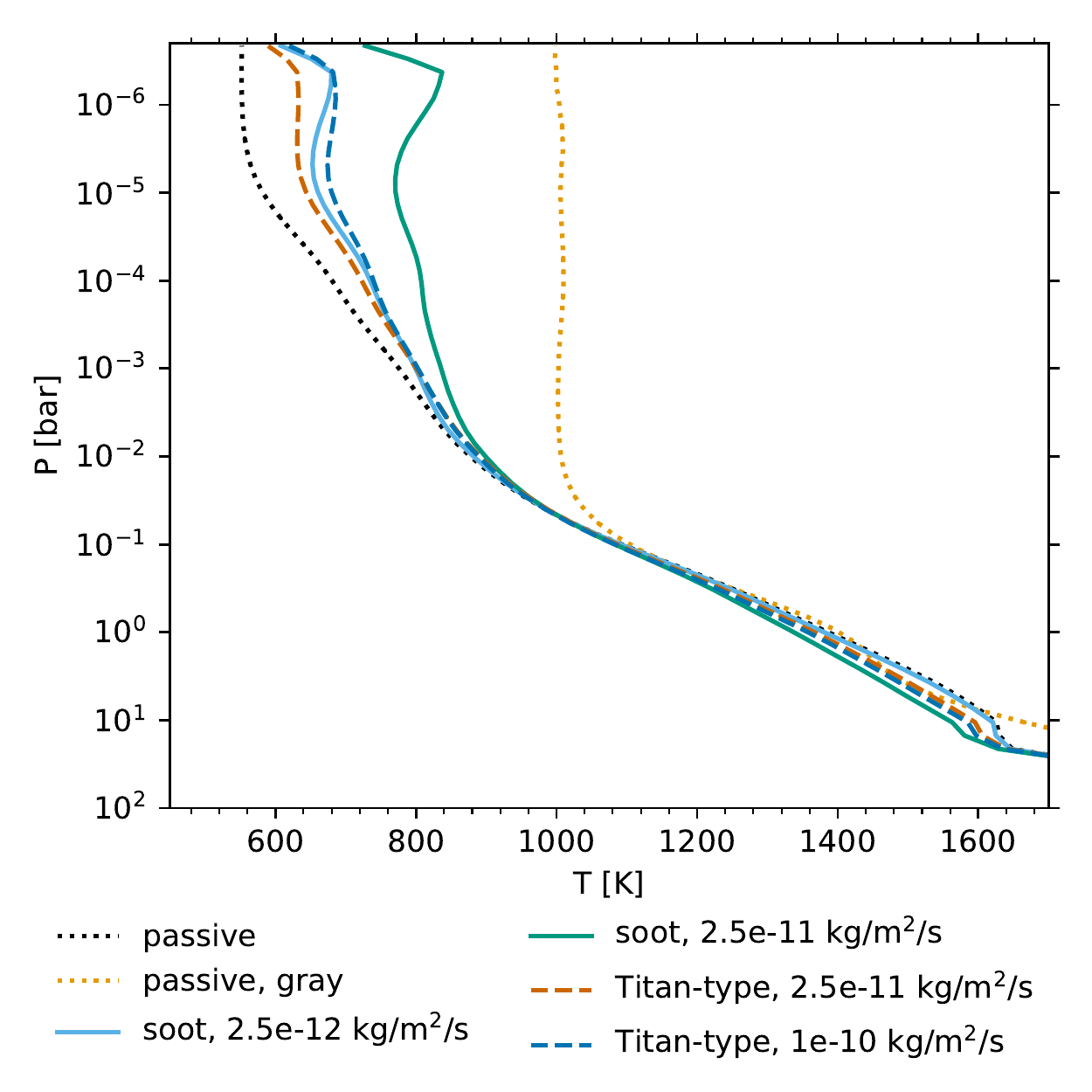}
\caption{Nightside average temperature profiles.}
\label{fig:temperatureaverages_night}
\end{center}
\end{figure}
Before looking at the effects of radiative feedback, we have to compare how the simulation results obtained from the model with wavelength-dependent, correlated-k radiative transfer without haze radiative feedback compare to the gray model used in \citet{SteinrueckEtAl2021}. The temperature structure differs substantially between the simulations (Fig. \ref{fig:temperatureaverages_day} and \ref{fig:temperatureaverages_night}). The gray simulation is almost isothermal for pressures $\lessapprox$10~mbar. In contrast, in the correlated-k simulation, the temperature declines steadily with decreasing pressure up until $\approx 10^{-5}$~bar. Below that pressure, the temperature profile becomes isothermal. Only for 10 bar$<p<$100~mbar, temperatures are similar. In this region, the nightside average temperatures are almost similar. The dayside average of the correlated-k model is somewhat cooler for $p\lessapprox1$~bar and somewhat hotter for $p\gtrapprox1$~bar. The double-gray model further significantly underestimates day-to-night temperature contrast for $p\lessapprox50$~mbar.
It is well-known that gray models overestimate temperatures at low pressures, both in 1D \citep[e.g.,][]{Guillot2010RadiativeEquilibrium} and 3D models \citep{LeeEtAl2021PicketFenceGCM}. This effect is particularly strong when choosing a constant-with-pressure opacity, as is the case in our double-gray model.

\begin{figure*}
\begin{center}
\plotone{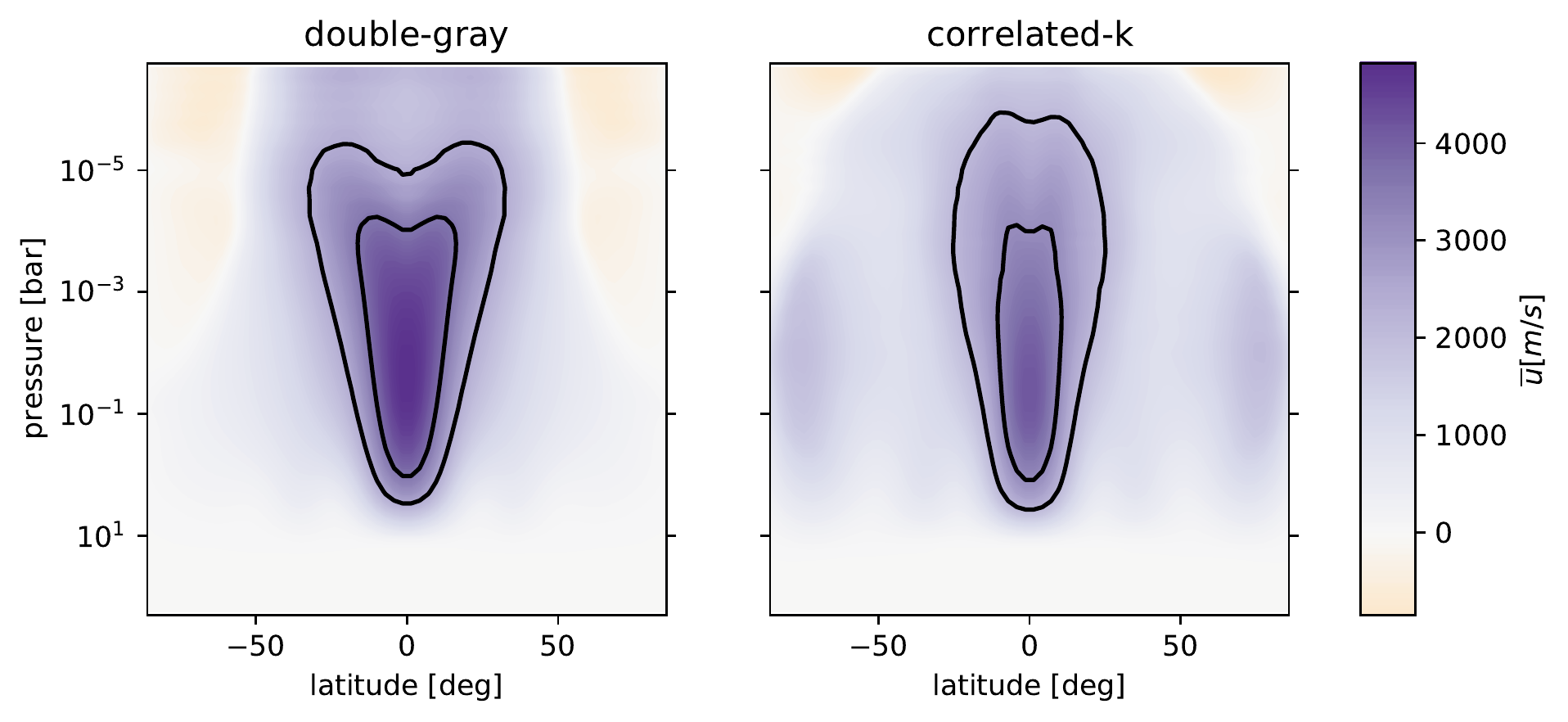}
\caption{Comparison of the zonal-mean zonal velocity in the double-gray (left panel) and correlated-k (right panel) simulations without haze radiative feedback. The contours outline the regions in which the zonal-mean zonal velocity is larger than 50\% and 75\% of its peak value within the simulation.}
\label{fig:zonalmeanvelocity}
\end{center}
\end{figure*}

\begin{figure*}
\begin{center}
\plotone{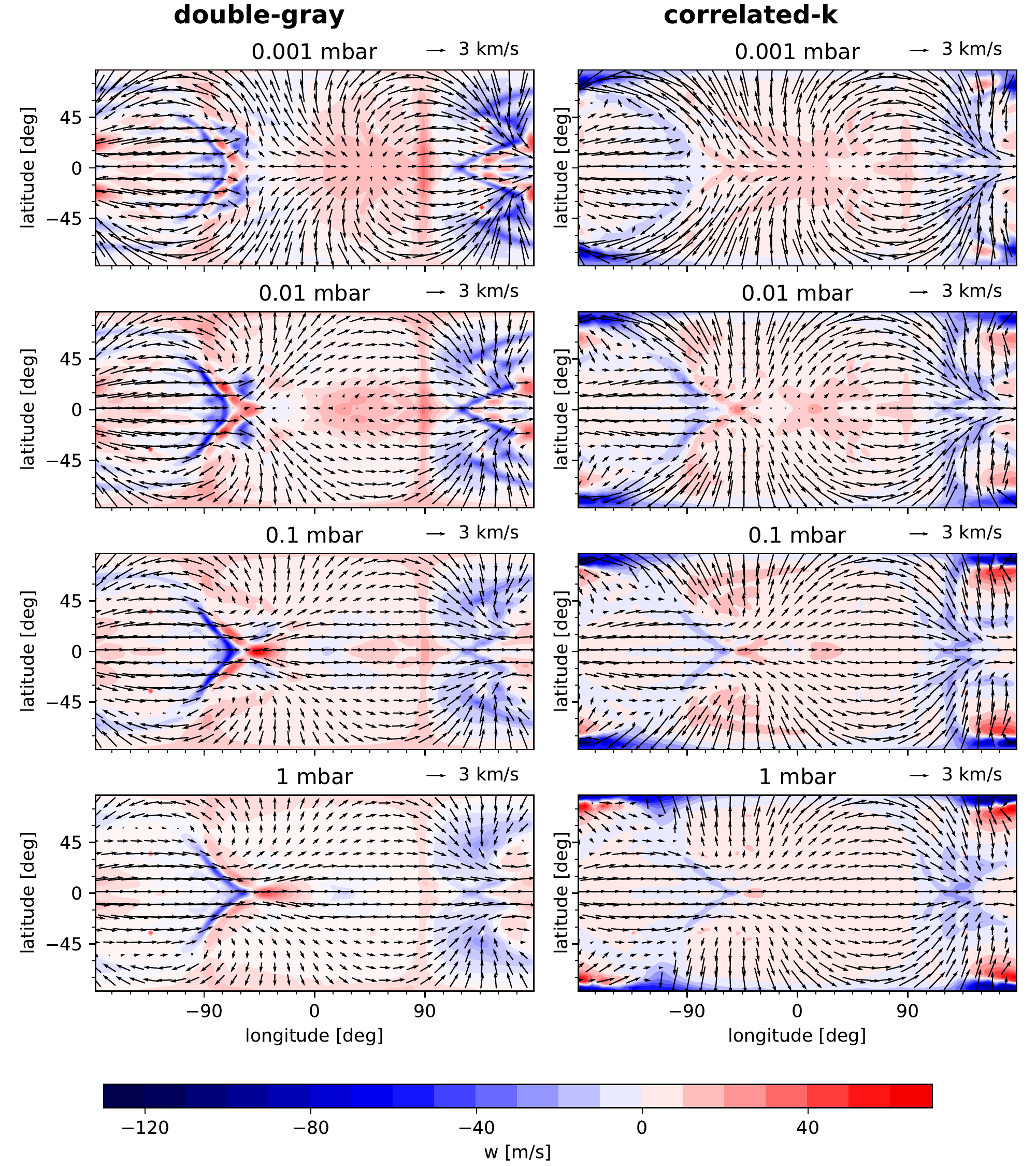}
\caption{Horizontal (arrows) and vertical (colorscale) velocities on isobars in the double-gray (left column) and correlated-k (right panel) simulations without haze radiative feedback. Positive vertical velocities correspond to upwelling. The substellar point is located at the center of each panel.}
\label{fig:velocities_grayvssparc}
\end{center}
\end{figure*}
Qualitatively, there are many similarities in the atmospheric circulation, including that both models produce predominantly day-to-night flow at low pressures and a strong super-rotating equatorial jet at higher pressures, typical for 3D simulations of hot Jupiters. Looking at the more detailed picture, however, there are significant differences.
A comparison of the zonal-mean zonal velocity is shown in Fig. \ref{fig:zonalmeanvelocity}. In the correlated-k simulation, the core region of the equatorial jet is more narrow in latitude than in the double-gray simulation. Further, in the double-gray simulation, the jet broadens with increasing altitude. In the correlated-k simulation, there is less broadening with altitude. Furthermore, the peak velocity drops from  $\approx4,800$~m~s$^{-1}$ in the gray simulation to $\approx4,200$~m~s$^{-1}$ in the correlated-k simulation. \citet{LeeEtAl2021PicketFenceGCM} also compared the changes in atmospheric circulation and temperature structure between a correlated-k and a double-gray model in a simulation of HD~209458b. Their findings are very similar to ours. The only exception to this is the peak strength of the equatorial jet, which in their model increases with the correlated-k approach, while it decreases in our model.

Looking at the horizontal velocities on isobars (shown in Fig. \ref{fig:velocities_grayvssparc} as arrows), perhaps the most striking change is that the location of the mid-latitude nightside vortices moves poleward and closer to the antistellar longitude in the correlated-k simulation. The shape of the vortices also becomes more asymmetrical. Further, there are significant changes in the vertical velocities. In the double-gray simulation, the largest vertical velocities are at the chevron-shaped morning terminator downwelling feature \citep[which previously has been identified as hydraulic jump, ][]{ShowmanEtAl2009, SteinrueckEtAl2021} and at mid-latitudes between evening terminator and substellar point. While there still is strong downwelling in these regions in the correlated-k simulation, the vertical velocities are somewhat lower than in the double-gray simulation. Instead, the largest downward vertical velocities are found on the nightside near the pole, at $\approx75^\circ$ latitude, near the antistellar longitude. At this location, downward velocities reach a value of 120 m~s$^{-1}$ at a pressure of 1 $\mu$bar. This is more than 1.5 times the peak vertical velocity at the same pressure level in the double-gray simulation. In both the double-gray and the correlated-k simulation, the regions of strong up- and downwelling remain vertically coherent for over three orders of magnitude (between 1 mbar and 1 $\mu$bar). We further note that in the double-gray simulation, there is a narrow band of strong upwelling at the evening terminator. In the correlated-k simulation, there are subtle hints of such a band but it is by far not as prominent.

\begin{figure*}
\begin{center}
\plotone{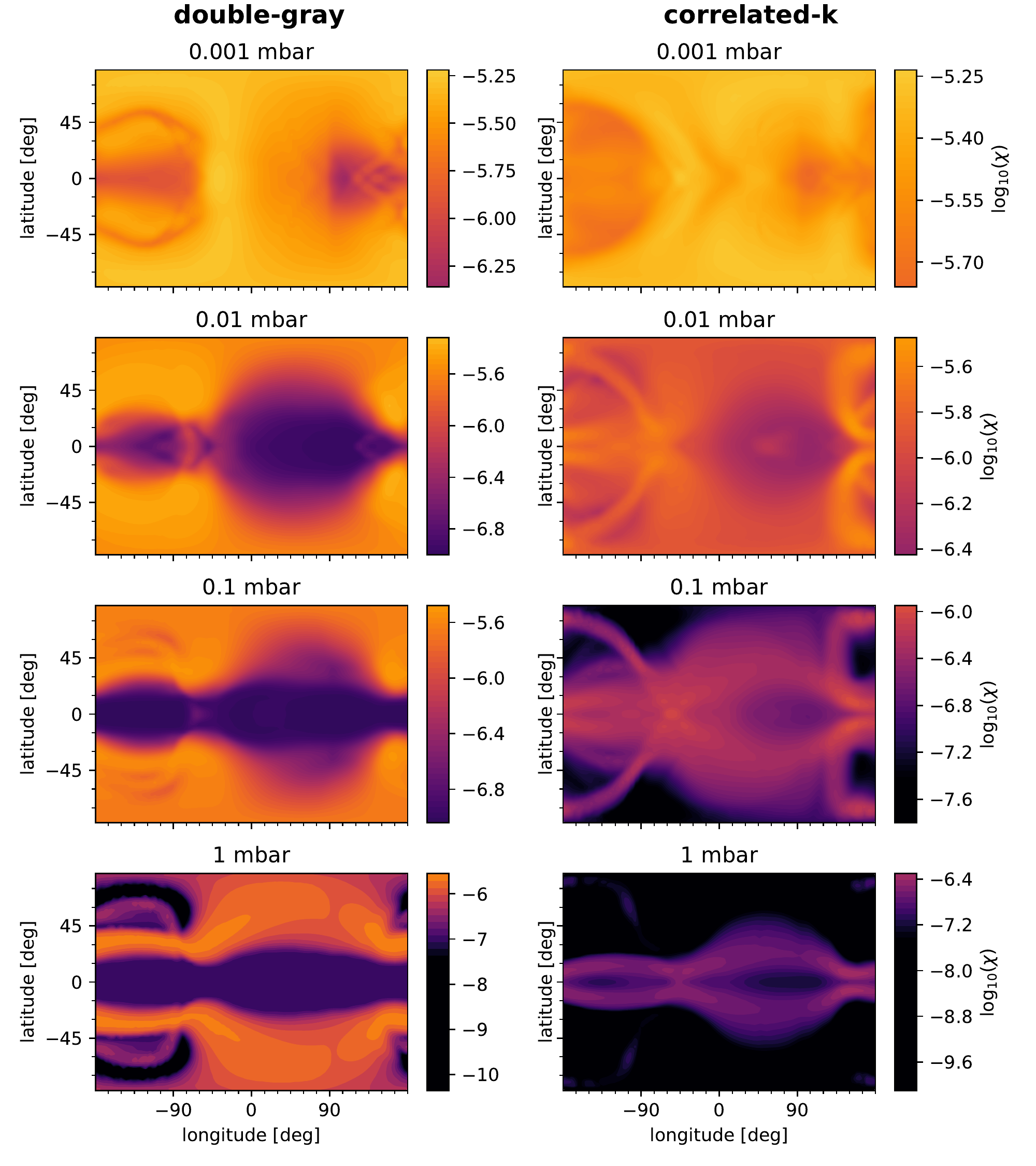}
\caption{Haze mass mixing ratio $\chi$ (on a logarithmic scale) on isobars in the double-gray (left column) and correlated-k (right panel) simulations without haze radiative feedback. The colorscale range has been chosen to be identical in both figures.}
\label{fig:mixingratioisobars_grayvssparc}
\end{center}
\end{figure*}
The differences in atmospheric circulation result in substantial changes in the three-dimensional haze distribution (Fig \ref{fig:mixingratioisobars_grayvssparc}). In the double-gray simulation, as described in detail in \citet{SteinrueckEtAl2021}, hazes accumulate in the mid-latitude nightside vortices between 3 $\mu$bar and 0.1 mbar. In the correlated-k simulation, instead, the haze mass mixing ratio remains low in the center of the vortices. However, there is a band of enhanced haze mass mixing ratio circling the center of the nightside vortices, following the horizontal projection of the streamlines. This band intersects with all three major downwelling regions (pole, west of antistellar point, near morning terminator). The haze mixing ratio clearly is further enhanced near these intersections. As the band almost reaches down to the equator, the equatorial region on the nightside also has enhanced haze mixing ratios. The equatorial region on the dayside and near the evening terminator, which is dominated by upwelling, is strongly depleted of hazes (especially east of the substellar point). At high latitudes on the dayside, there are intermediate mixing ratios. As pressure increases, the mixing ratio on most of the dayside decreases only slowly, while the mixing ratio in the enhanced regions decreases much faster. Thus, the circular bands with enhanced mixing ratio surrounding the nightside vortices lose their prominence with increasing pressure. The horizontal haze distribution thus gradually morphs into a pattern that resembles two broad bands of enhanced haze mixing ratio spanning around the planet, broadening and moving to higher latitudes on the dayside. This pattern qualitatively resembles the banded pattern at pressures above 0.1~mbar in the double-gray simulation \citep{SteinrueckEtAl2021}. The bands, however, are closer to the equator in the correlated-k simulation and both bands connect at the equator near the morning terminator.

\begin{figure}
\begin{center}
\includegraphics[width=\columnwidth]{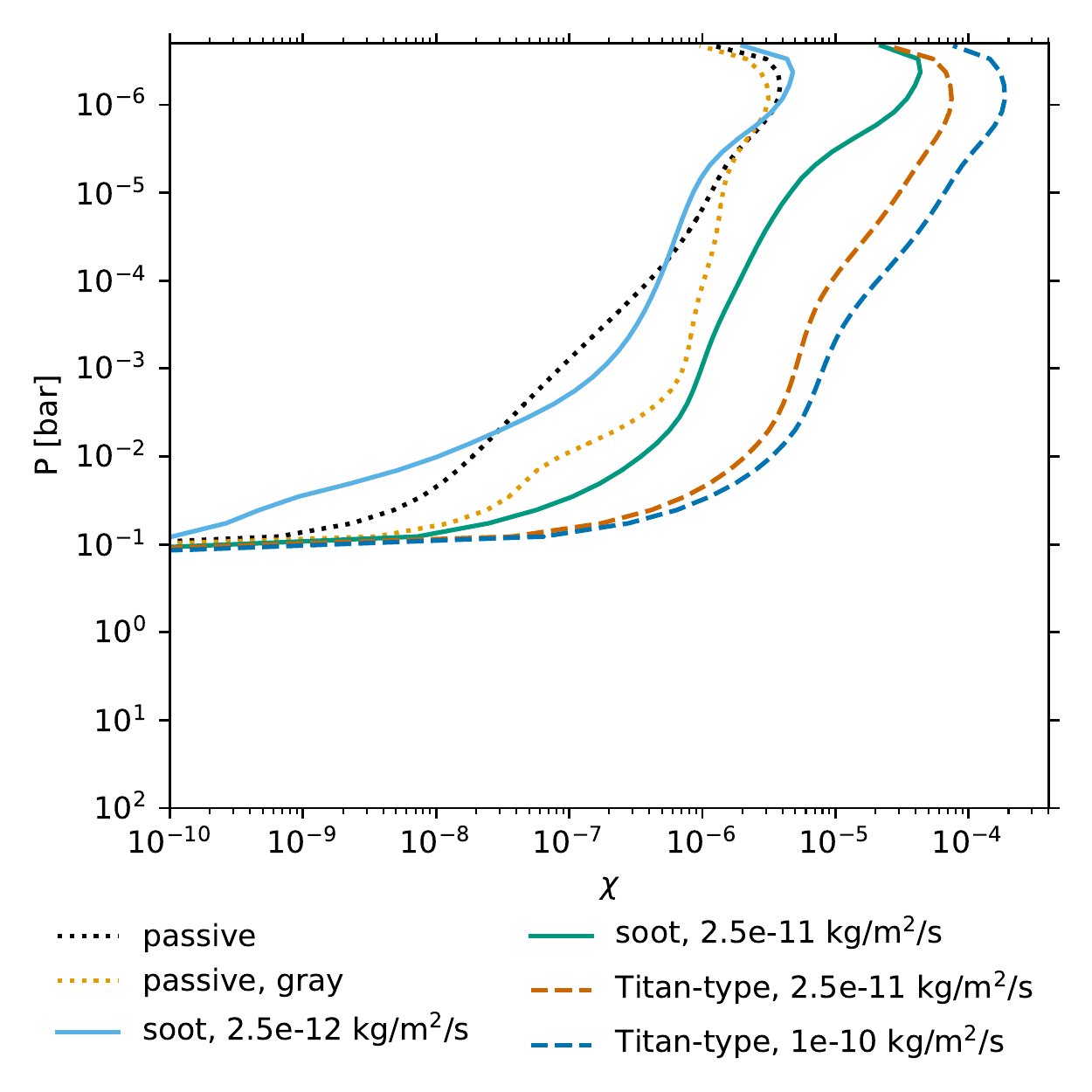}
\caption{Global average profiles of the haze mass mixing ratio.}
\label{fig:mmraverages_global}
\end{center}
\end{figure}
Comparing the globally-averaged vertical profiles of the haze mass mixing ratio (Fig. \ref{fig:mmraverages_global}), the mixing ratio drops off much faster with increasing pressure in the correlated-k simulation. Presumably, this can be attributed to the stronger downwelling velocities. In addition, the mass mixing ratio gradient remains more constant with pressure in the correlated-k simulation. These changes have implications for the transmission spectrum when comparing simulations with the same haze distribution rates, as discussed in Section \ref{subsec:transitspectra}.

\section{Simulations with haze radiative feedback}
\label{sec:hazefeedback}
\subsection{Soot-like refractive index\label{subsec:soothazes}}
In the simulations with haze radiative feedback and soot-like refractive index, the dayside temperature increases dramatically at low pressures compared to the simulation without haze feedback (Fig. \ref{fig:temperatureaverages_day}). At the 1~$\mu$bar level, near the center of the haze production region, the change is as high as 700~K in the simulation with the highest haze production rate ($2.5\cdot10^{-11}$~kg~m$^{-2}$~s$^{-1}$) and 400~K in the simulation with the lowest haze production rate ($2.5\cdot10^{-12}$~kg~m$^{-2}$~s$^{-1}$). On the nightside (Fig. \ref{fig:temperatureaverages_night}), in contrast, the temperature increase is quite moderate. This means that the day-to-night temperatures contrast increases significantly at pressures $<10$~mbar, from about 200~K to 400~K in the simulation with the lowest haze production rate and 500 to 700~K (depending on pressure) in the simulation with the highest haze production rate.

In the dayside-averaged temperature profile (Fig. \ref{fig:temperatureaverages_day}), two distinct thermal inversions are present that are separated by a temperature minimum near 10~$\mu$bar, just below the haze production region. This temperature minimum is not observed in 1D simulations and is a result of the interaction of hazes with atmospheric dynamics (as further explained towards the end of this section).

\begin{figure*}
\begin{center}
\includegraphics[width=\textwidth]{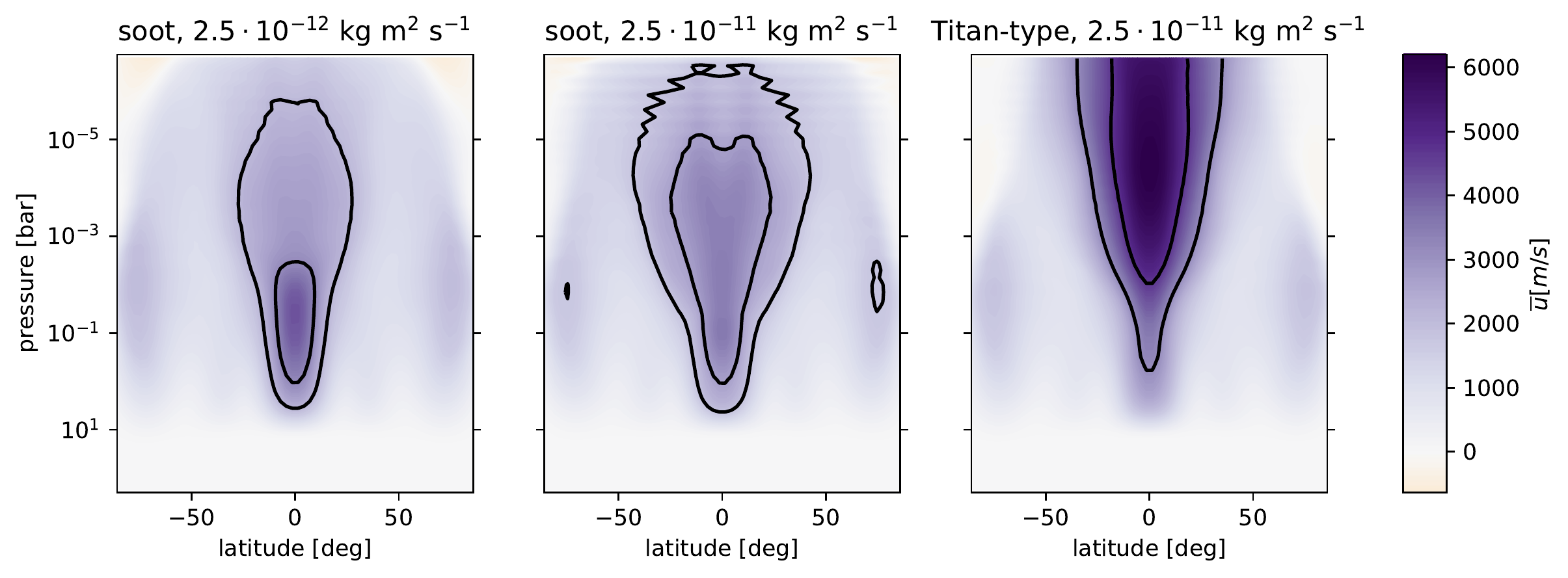}
\caption{Comparison of the zonal-mean zonal velocity in simulations with haze radiative feedback. The contours outline the regions in which the zonal-mean zonal velocity is larger than 50\% and 75\% of its peak value within the simulation.}
\label{fig:zonalmeanvelocity_hazefeedback}
\end{center}
\end{figure*}

\begin{figure*}
\begin{center}
\plotone{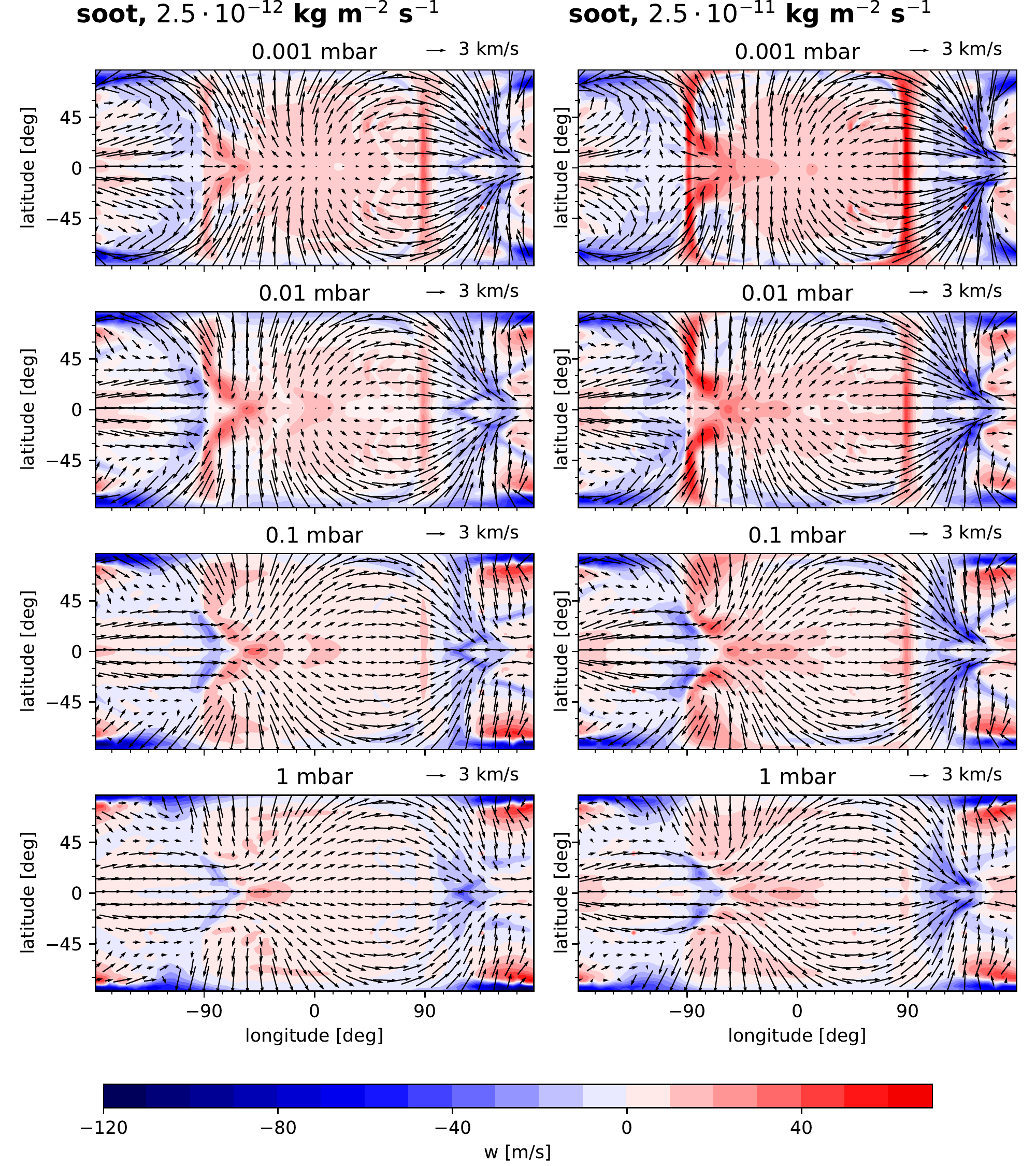}
\caption{Horizontal (arrows) and vertical (colorscale) velocities on isobars in simulations with haze radiative feedback using soot refractive indices with two different haze production rates (left column: $2.5\cdot10^{-12}$~kg~m$^{-2}$~s$^{-1}$, right colum: $2.5\cdot10^{-11}$~kg~m$^{-2}$~s$^{-1}$). Positive vertical velocities correspond to upwelling. The substellar point is located at the center of each panel.}
\label{fig:velocities_soot}
\end{center}
\end{figure*}
The haze radiative feedback significantly alters atmospheric circulation. Looking at the zonal-mean zonal velocity, the equatorial jet broadens significantly in latitude while its overall strength decreases (Fig. \ref{fig:zonalmeanvelocity_hazefeedback}). The strength of upwelling on the dayside increases substantially (Fig. \ref{fig:velocities_soot}). In particular, the narrow upwelling region at the evening terminator that appeared in the double-gray simulation but was barely visible in the correlated-k simulation without haze radiative feedback appears again and becomes much stronger for increased haze production rates. The chevron-shaped downwelling and adjacent upwelling feature at the morning terminator associated with the hydraulic jump significantly changes its shape as well. While a chevron-shape is retained close to the equator, additional upwelling parallel to the terminator appears at higher latitudes. The downwelling regions on the nightside become less localized. Their peak velocity is reduced significantly, but downwelling is distributed over a much larger region. In a very rough sense, one could say that the atmospheric circulation with soot-based haze radiative feedback changes in a way that makes it more similar to the double-gray simulation, especially for the cases with low-to-intermediate haze production rates. This is likely because the absorption cross section of soot has a relatively weak and smooth wavelength dependence. Therefore, adding soot opacity at low pressures somewhat resembles adding a gray opacity at these regions.

\begin{figure*}
\begin{center}
\plotone{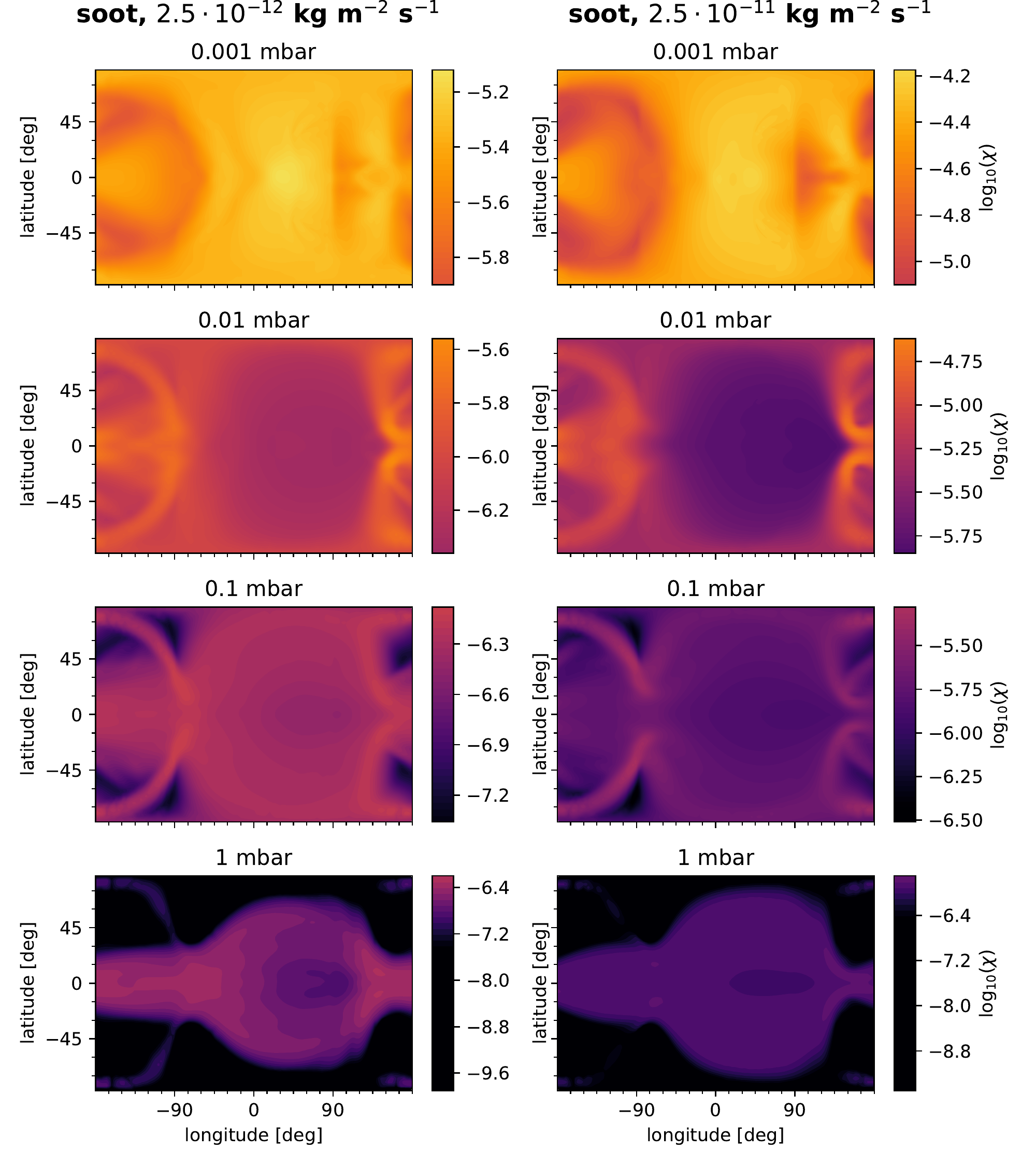}
\caption{Haze mass mixing ratio $\chi$ (on a logarithmic scale) on isobars in simulations with haze radiative feedback using soot refractive indices with two different haze production rates (left column: $2.5\cdot10^{-12}$~kg~m$^{-2}$~s$^{-1}$, right colum: $2.5\cdot10^{-11}$~kg~m$^{-2}$~s$^{-1}$). The colorscale range has been chosen to be identical to the one in Fig. \ref{fig:mixingratioisobars_grayvssparc} in the left column, while it has been offset by a factor of 10 in the right column to facilitate the comparison between different haze production rates.}
\label{fig:mixingratioisobars_soot}
\end{center}
\end{figure*}
In general, the horizontal distribution of the hazes (Fig. \ref{fig:mixingratioisobars_soot}) remains qualitatively similar to the distribution in the passive correlated-k simulation. The center of the nightside vortices remains depleted of hazes. Again, below the haze production region, there are bands of enhanced haze mixing ratio surrounding the center of the vortices, with localized higher haze mixing ratios where the bands intersect with the downwelling areas.  At somewhat higher pressures ($p\gtrapprox0.1$~mbar), the dayside haze mixing ratio becomes more uniform compared to the simulation without haze radiative feedback and the equatorial region is no longer depleted. Thus, rather than having one narrower circumplanetary band with increased haze mixing ratio in each hemisphere, there is one broader band that includes the equatorial region.

\begin{figure}
\begin{center}
\includegraphics[width=\columnwidth]{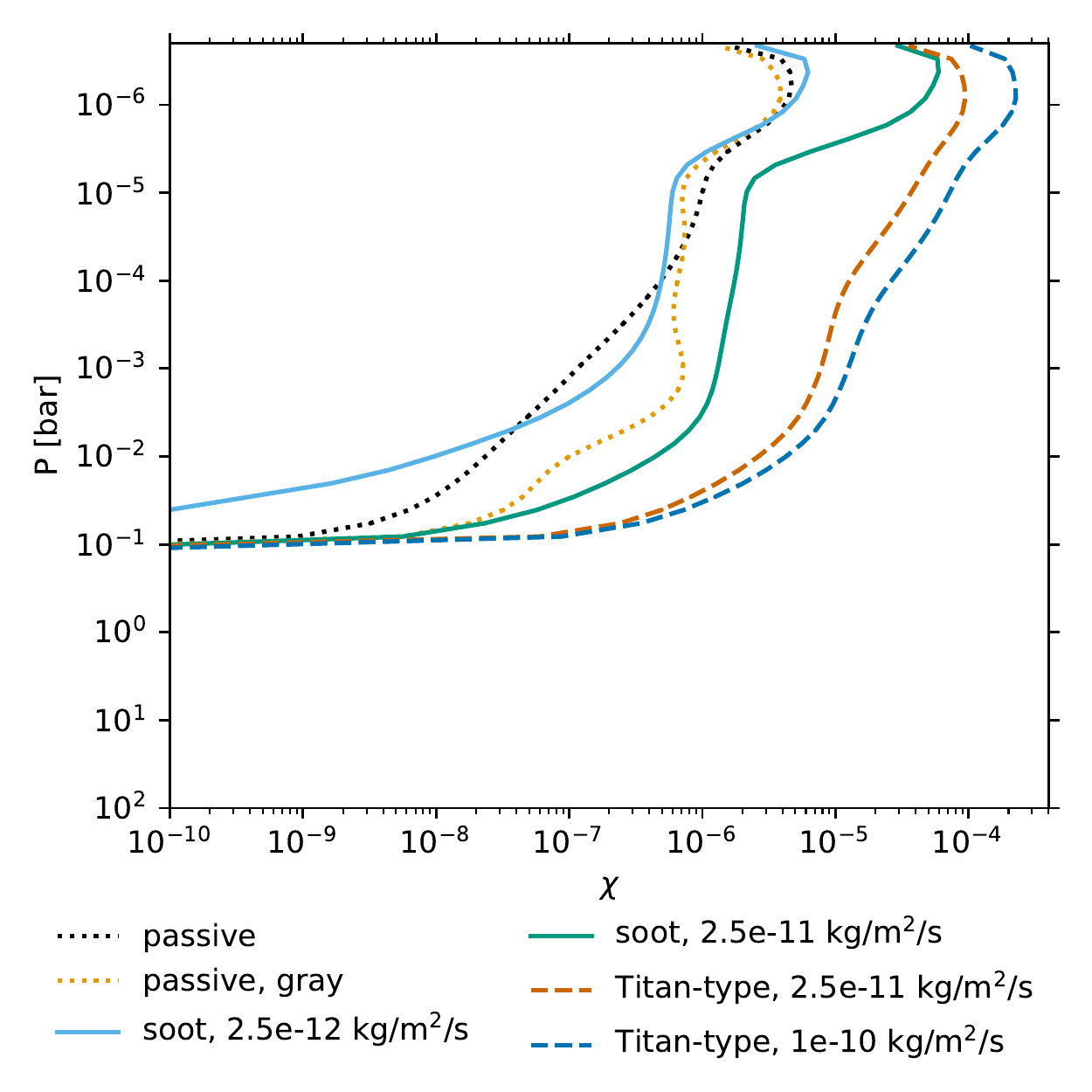}
\caption{Dayside average profiles of the haze mass mixing ratio, calculated using an average weighted by the cosine of the angle of incidence.}
\label{fig:mmraverages_day}
\end{center}
\end{figure}
The globally-averaged vertical mixing ratio profiles (Fig. \ref{fig:mmraverages_global}) qualitatively change compared to the wavelength-dependent, passive  simulation: The mass mixing ratio declines less evenly (like in the passive, gray case). 
In addition, it is insightful to also examine the dayside-averaged haze mixing ratio profiles (Fig. \ref{fig:mmraverages_day}). One can see that for the soot haze radiative feedback simulations, the mixing ratio profile on the dayside is close to constant for a significant pressure region. This is because of the stronger upwelling on large portions of the dayside. The extent of the pressure region with almost-constant mixing ratio increases with a higher haze production rate. Likely, this is partially caused by the increased upward velocities on the dayside. However, the fact that the equatorial jet further weakens in the $2.5\cdot10^{-11}$~kg~m$^{-2}$~s$^{-1}$ simulation could also contribute, as the jet acts to homogenize the mixing ratio between day- and nightside.
 
\begin{figure}
\begin{center}
\includegraphics[width=\columnwidth]{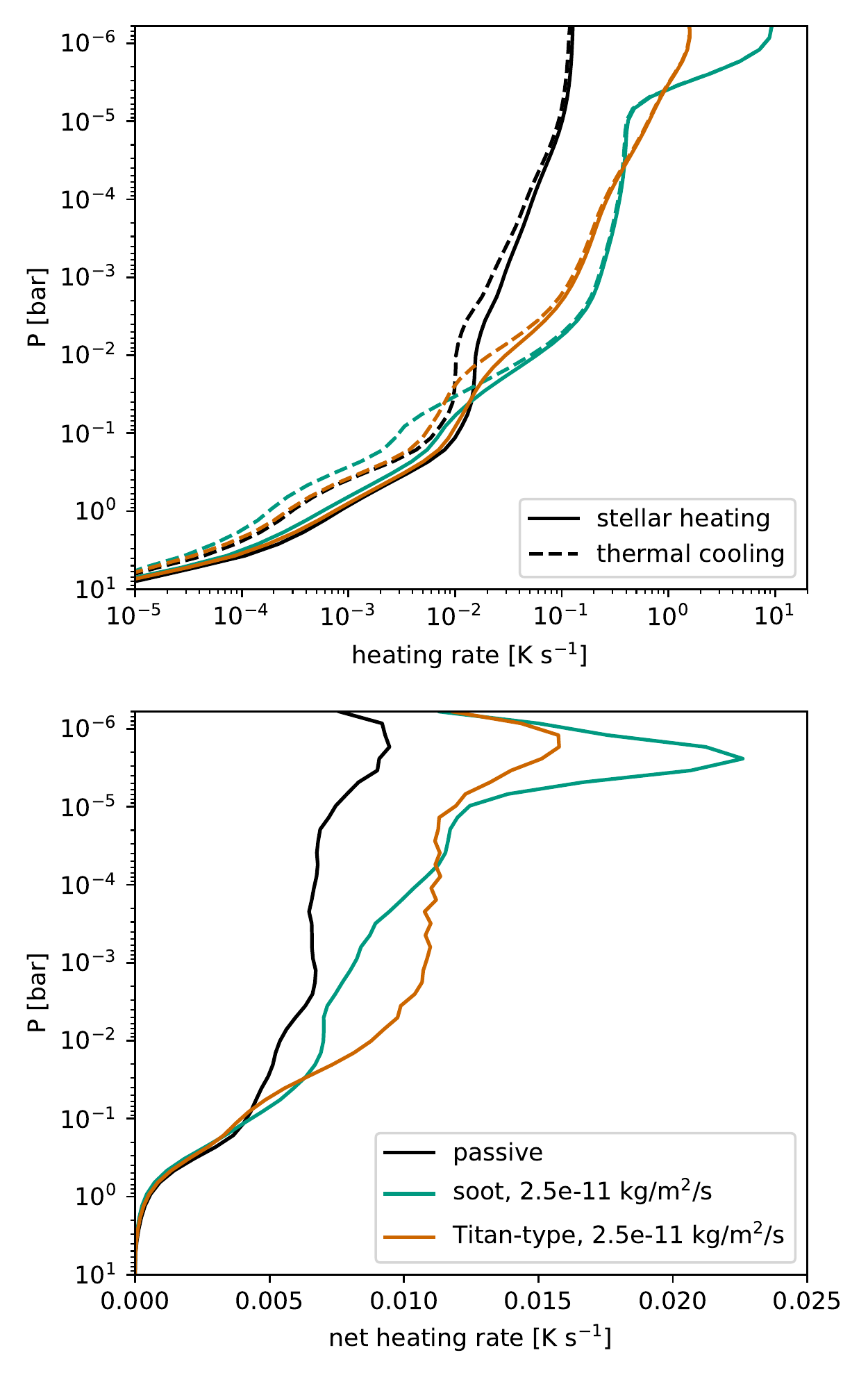}
\caption{Dayside-average profiles of the radiative heating rates in the passive simulation as well as two simulations with identical haze production rates, calculated using an average weighted by the cosine of the angle of incidence. The top panel shows the stellar heating and thermal cooling rates on a logarithmic scale, while the bottom panel shows the net difference between stellar heating and thermal cooling on a linear scale. The heating rates shown are instantaneous from simulation snapshots at 4,500~days, i.e., unlike other quantities shown in this work, they are not time-averaged. The top three layers of the model have been omitted due to boundary effects.}
\label{fig:heatingrates_day}
\end{center}
\end{figure}

The vertically almost homogeneous haze mixing ratio on the dayside  below the haze production region is also directly tied to the temperature minimum below the haze production region near 10~$\mu$bar. Upwelling on the dayside transports air with relatively low haze mixing ratio upwards from deeper layers, causing relatively low mass mixing ratios and thus low rates of stellar heating  (Fig. \ref{fig:heatingrates_day}, top panel) just below the haze production region. In the haze production region, the mass mixing ratio (and thus also the heating rate) then increases much faster with height than seen in the global average or in one-dimensional models which assume mixing to act only in a diffusive way.

\subsection{Titan-type refractive index\label{subsec:titanhazes}}
\begin{figure*}
\begin{center}
\plotone{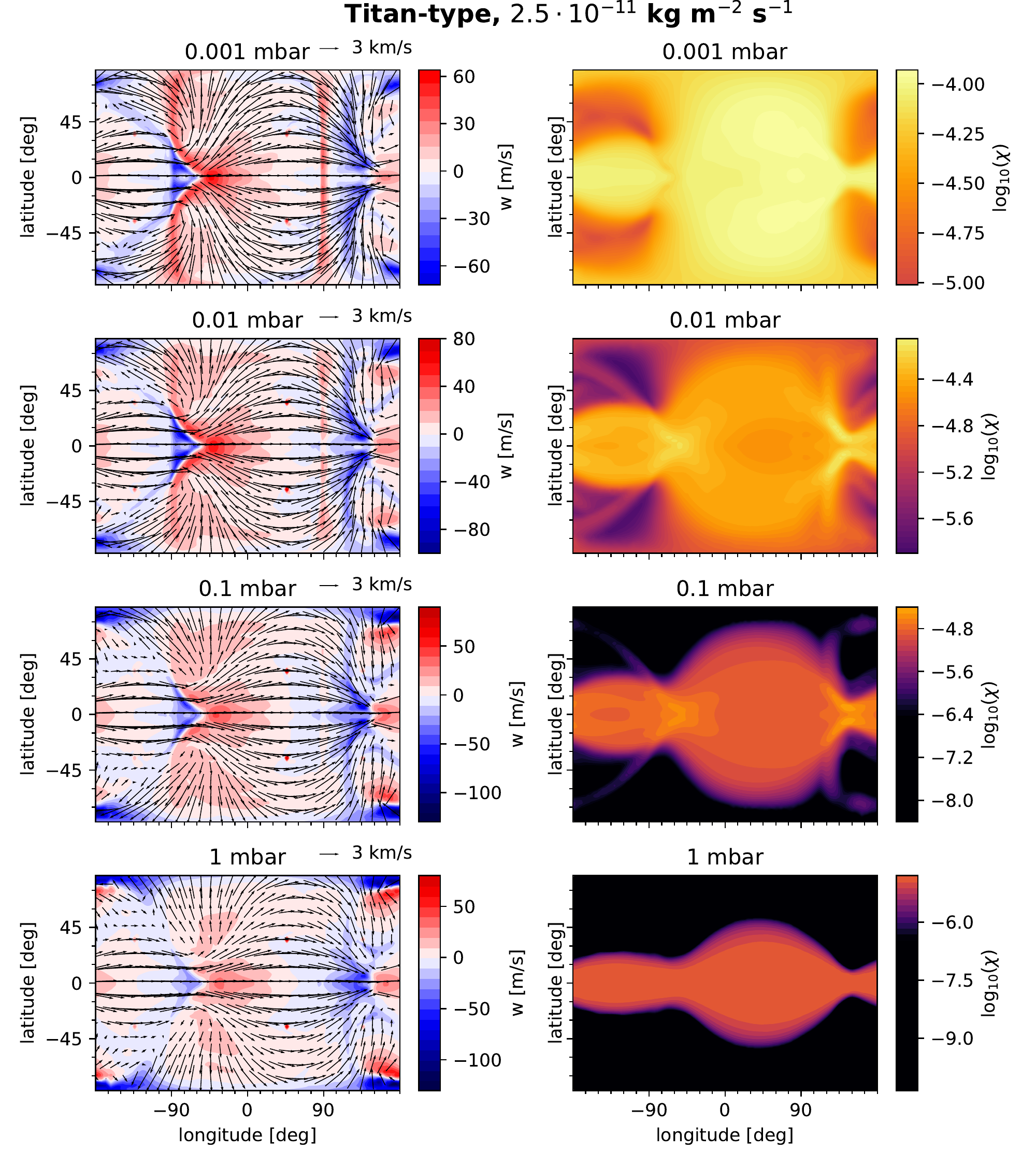}
\caption{Left column: Horizontal (arrows) and vertical (colorscale) velocities on isobars in a simulations with haze radiative feedback using Titan-type refractive indices. Positive vertical velocities correspond to upwelling. The substellar point is located at the center of each panel. Right column: Haze mass mixing ratio $\chi$ (on a logarithmic scale) on isobars in simulations with haze radiative feedback using soot refractive indices with two different haze production rates (left column: $2.5\cdot10^{-12}$~kg~m$^{-2}$~s$^{-1}$, right colum: $2.5\cdot10^{-11}$~kg~m$^{-2}$~s$^{-1}$). Compared to Fig. \ref{fig:mixingratioisobars_grayvssparc}, the colorscale has been offset by a factor of 10 in the right column to facilitate the comparison between different haze production rates.}
\label{fig:velocities_tholin}
\end{center}
\end{figure*}

Compared to any of the other simulations, the atmospheric circulation changes dramatically in the simulations with Titan-type hazes (Fig. \ref{fig:velocities_tholin}). The strength of the equatorial jet increases drastically, especially at low pressures (Fig. \ref{fig:zonalmeanvelocity_hazefeedback}). While in all other simulations, there is westward flow on at least parts of the dayside, especially west of the substellar point, close to the peak of the haze production profile (2~$\mu$bar), in these two simulations, there is eastward flow throughout the entire dayside. This substantially changes the 3D distribution of the hazes (Fig. \ref{fig:velocities_tholin}). In the haze production region, hazes are now advected eastward from the dayside, resulting in a higher haze mixing ratio at the evening terminator than at the morning terminator. This is the opposite of what was observed in all of the simulations with passive tracers (both in the double-gray and the correlated-k case) and radiative feedback with soot refractive indices. Below the haze production region, the equatorial jet (which widens substantially on the dayside) homogenizes haze abundances across the equatorial region and most of the dayside. The only regions that remains depleted of hazes are the night side vortices. Even deeper in the atmosphere ($\approx1$~mbar), the hazes remain mostly in the equatorial region. 

We further note that the dayside-average temperature (Fig. \ref{fig:temperatureaverages_day}) and mass mixing ratio profiles (Fig. \ref{fig:mmraverages_day}) in the Titan-type case also are qualitatively distinct from the soot-like case. In the case with a haze production rate of $2.5\cdot10^{-11}$~kg~m$^{-2}$~s$^{-1}$, the temperature profile is isothermal below the haze production region (compared to the double-peaked thermal inversion and the local temperature minimum below the haze production region in the soot-like case). For the haze production rate of $1\cdot10^{-10}$~kg~m$^{-2}$~s$^{-1}$, the temperature profile decreases with increasing pressure below the haze production rate. The dayside-average haze mass mixing ratio steadily decreases with increasing pressure, closely resembling the globally-averaged haze mass mixing ratio profile (Fig. \ref{fig:mmraverages_global}). There is no region of almost-constant haze mass mixing ratio below the haze production region (as seen in the soot simulations). Presumably, these changes are directly linked to the fact that the hazes are now efficiently transported between day-and nightside in the equatorial region. Finally, the globally-average haze mass mixing ratio is noticeably larger than in the soot-like case with the same haze production rate.

To examine possible causes for the differences in atmospheric circulation between soot-like and Titan-type hazes, we calculated the instantaneous heating rates at 4,500 days simulation time. Figure \ref{fig:heatingrates_day} shows the dayside average of the radiative heating rates for the two simulations with the same haze production rate ($2.5\cdot10^{-11}$~kg~m$^{-2}$~s$^{-1}$), as well as in the passive simulation. The stellar heating profiles (top panel) differ dramatically between soot and Titan-type hazes. In the soot simulation, the stellar heating is highly concentrated near the peak of the haze production region and rapidly drops off until below the haze production region and then remains relatively constant between 10 $\mu$bar and 1 mbar. In contrast, in the Titan-type simulation, the heating rate peaks at a much lower value in the haze production region and declines more gradually with increasing pressure.

For atmospheric dynamics, the most relevant quantity is the net radiative heating rate (bottom panel).  Both simulations with haze feedback exhibit overall higher net heating rates at pressures below 100~mbar than the passive simulation, with a large peak in the haze production region. However, in the simulation with Titan-type haze, the peak value is only 2/3 of the peak value in the soot simulation. At the same time, there is more heating in the pressure region between 0.1~mbar and 30~mbar in the Titan-type simulation. Thus, radiative heating is spread out over a larger pressure range in the Titan-type simulation, while it is concentrated at low pressures in the soot case. This is expected, as the extinction cross-section of Titan-type hazes has a much larger wavelength dependence, meaning that in some wavelength regions, the radiation can penetrate much deeper into the atmosphere than at other wavelengths. We suggest that the additional energy deposition between 0.1~mbar and 30~mbar drives the stronger and vertically more extended equatorial jet in the Titan-type case. In contrast, in the soot-like case, the additional energy deposited directly in the haze production region likely cannot drive the equatorial jet because at pressures this low, the radiative timescale is much shorter than the wave propagation timescale and thus the dynamic mechanism for driving the equatorial jet is inhibited \citep{PerezBeckerShowman2013,KomacekShowman2016}.

\section{Predicted observations}
\label{sec:observations}
\subsection{Transmission spectra \label{subsec:transitspectra}}
\begin{figure*}
\begin{center}
\plotone{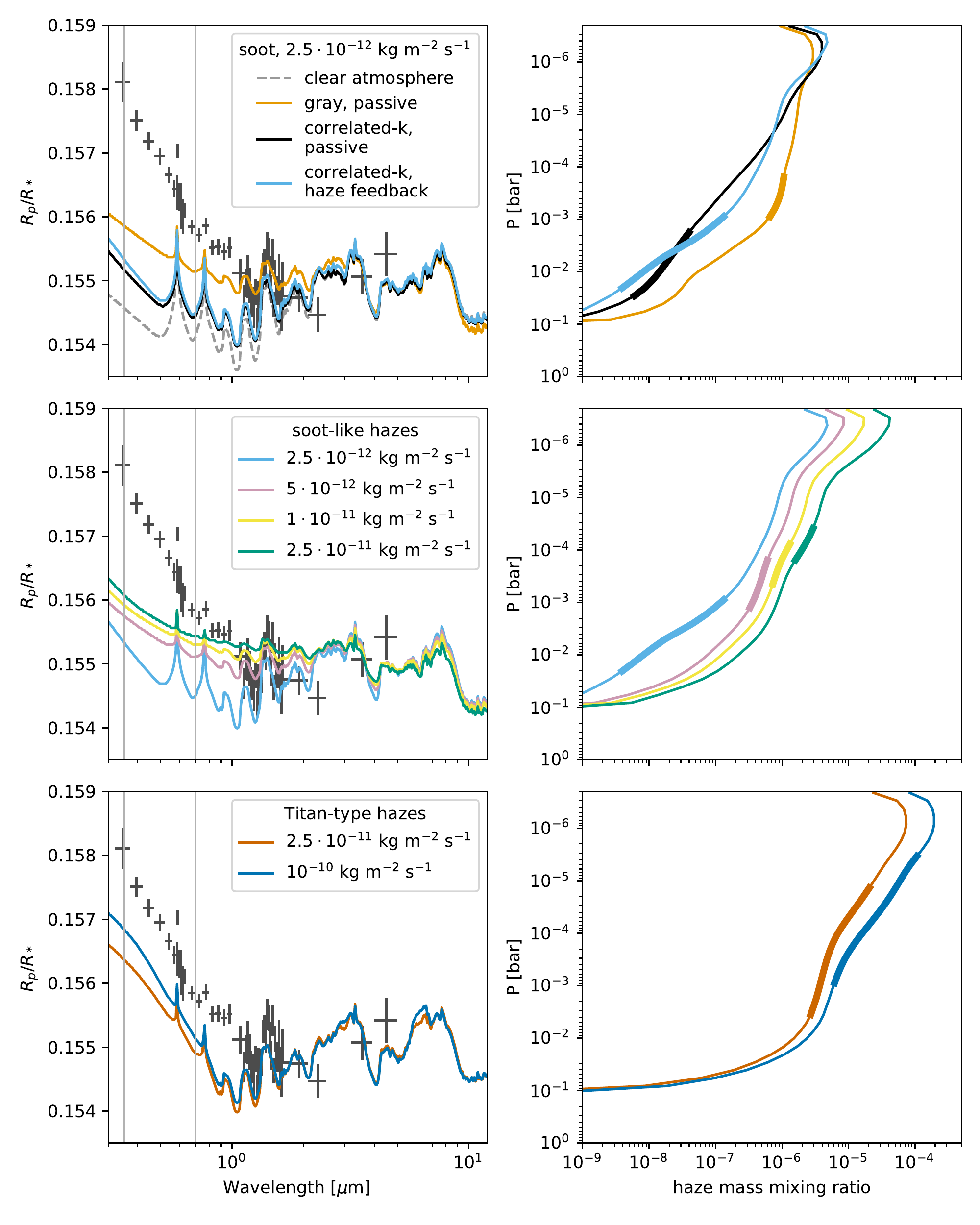}
\caption{Left column: Model-predicted transmission spectra for different simulations. Black crosses represent observational data \citep{PontEtAl2013,McCulloughEtAl2014} using the analysis by \citep{SingEtAl2016}. The gray vertical lines indicate wavelengths of 350 and 700~nm, corresponding to the pressure regions highlighted in the right column. Right column: Mass mixing ratio averaged across the terminator region for the simulations shown in the left column. For each simulation, the pressure region probed by continuum extinction between 350 and 700~nm (i.e. not including pressures probed by the core of the sodium line) is highlighted as thick line. }
\label{fig:transitspectra}
\end{center}
\end{figure*}

First, we compare the wavelength-dependent models without and with haze radiative feedback to the double-gray model with the best-fit haze production rate  while keeping the haze production rate constant (Fig. \ref{fig:transitspectra}, panel (a)). At the same haze production rate, the wavelength-dependent models (with and without haze radiative feedback) show stronger near-infrared features as well as a steeper short-wavelength slope. The short-wavelength slope is almost parallel to the observed slope. However, there is a large offset, with $R_p/R_s$ being about 0.002 lower in the models compared to the observations. The steeper slope is consistent with the haze mass mixing ratio declining faster with increasing pressure in the simulations with wavelength-dependent radiative transfer. This decline results in a lower haze mass mixing ratio below the haze production region (i.e. at pressures higher than $10^{-5}$~bar), which explains the larger near-infrared features. 

Compared to the simulation with wavelength-dependent radiative transfer and passive hazes, the simulation with haze radiative feedback has an even steeper short-wavelength slope. This can largely be attributed to a stronger mixing ratio gradient in the pressure region probed in the simulation with haze feedback (panel (b)). The higher temperature in the haze feedback simulation also may contribute, however, due to the low haze mass mixing ratio in both simulations, the transit spectrum is probing relatively deep in the atmosphere in the near-infrared (ca. 1-100 mbar). The temperature difference between both simulations is much smaller at these pressures compared to higher altitudes.

Given the stronger near-infrared features when keeping the haze production rate constant, it is necessary to look at simulations with increased haze production rates to assess whether radiative feedback can improve the match to observations. Panel (c) of Fig. \ref{fig:transitspectra} therefore shows transmission spectra from simulations with radiative feedback of soot-like hazes for different haze production rates. The best match to the WFC3 data is produced by the simulation with a haze production rate of $5\cdot 10^{-12}$ kg~m$^{-2}$~s$^{-1}$, twice as large as the haze production rate of the best-fit double-gray model. As the haze production rate increases, the short-wavelength slope in general becomes shallower. The reason is that with increased haze opacity, lower pressures with a weaker mass mixing ratio gradient are probed (panel (d)). In addition, the shape of the vertical mixing ratio profile at the terminator changes with increasing haze production rate, especially between 1 and 100~mbar, such that the mixing ratio gradient is less constant with pressure. None of the models with soot hazes thus match the observed transmission spectrum at short wavelengths.

Titan-type hazes absorb much less in the infrared, therefore higher haze production rates are needed to match near-infrared spectra. For both haze production rates simulated, the short-wavelength slope is steeper than in all soot-like models and is roughly parallel to the observed slope. The reason for the steeper slope is the extinction coefficient dropping off by about two orders of magnitude between the UV and the near-infrared. This is well-known and and also has been noted in previous work \citep[e.g.,][]{OhnoKawashima2020,SteinrueckEtAl2021,LavvasArfaux2021}. Overall, the Titan-type spectra match the transit observations better. However, a smaller but substantial offset between the short-wavelength observations and the models remains.

Overall, the Titan-type hazes can reproduce two of the main features of the measured transmission spectrum fairly well: the optical slope and the strengths of the near-infrared H$_2$O absorption feature. The main disagreement is the absolute transit depth in the optical, which differs by a few hundred parts per million. However, such a discrepancy could potentially arise from stellar activity; HD 189733 is known to be active \citep{Boisse2009}. Another possibility is that the absolute level of the measured spectrum is biased by visit-long time-dependent systematic correction \citep{Stevenson2014}. As \citet{ArfauxLavvas2022} note, an offset between the optical and near-infrared data due to either of these effects could also bring the \textit{HST} optical spectrum into agreement with the transit depth observed by \textit{SOFIA} \citep{AngerhausenEtAl2015SOFIA}. Given these possible systematic uncertainties in the transit depth measured with different instruments, the GCM results with Titan-type hazes agree fairly well with the overall morphology of the spectrum.

\begin{figure}
\begin{center}
\includegraphics[width=\columnwidth]{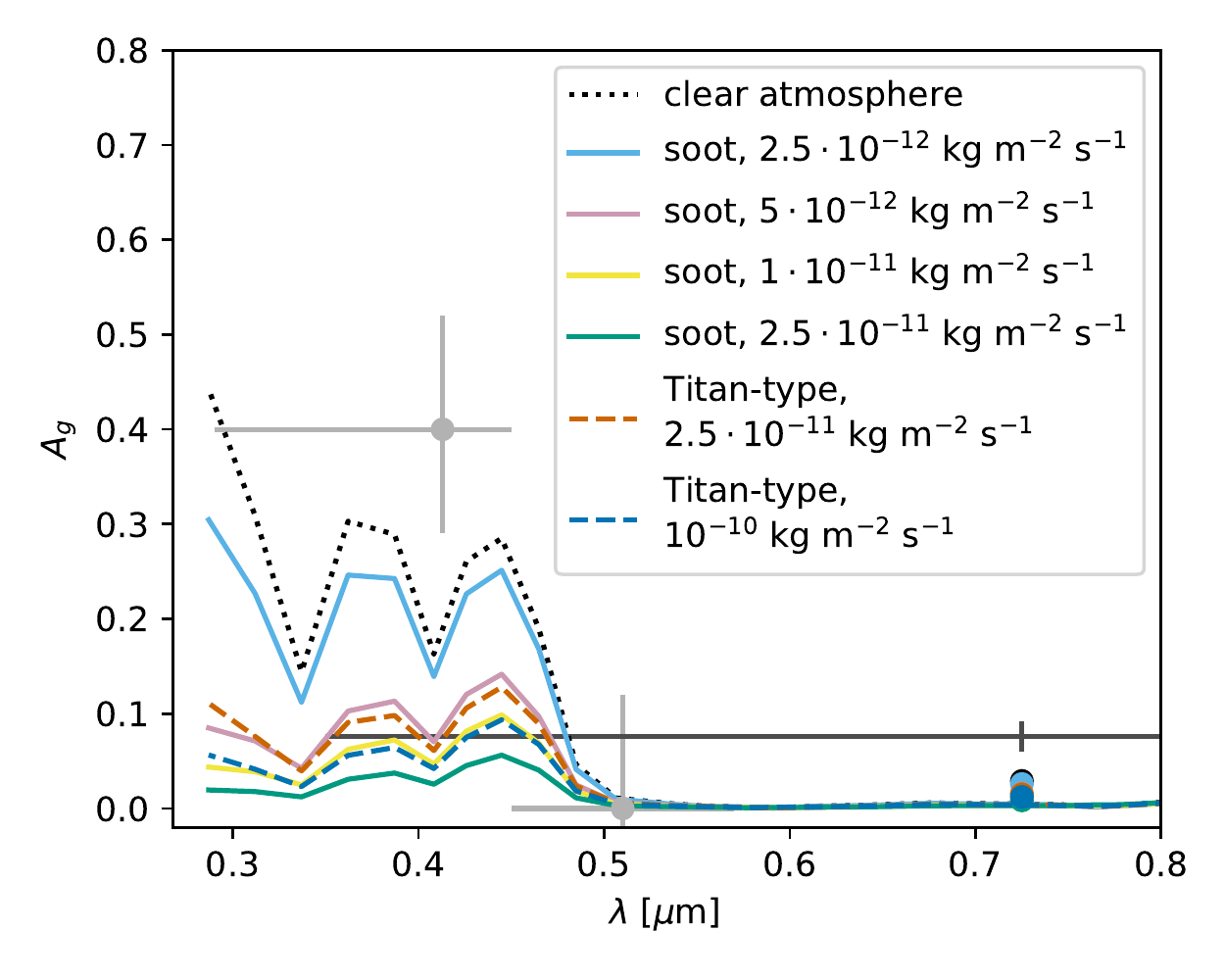}
\caption{Geometric albedo predicted from our models, together with observations of HD~189733b. Data points from \textit{HST STIS} with the G430L grating \citep{EvansEtAl2013} are shown in light gray. The measurement in the CHEOPS bandpass \citep{KrennEtAl2023HD189CHEOPS} is shown in dark gray, along with the model spectra integrated across the CHEOPS bandpass as filled circles. Note that the CHEOPS bandpass extends to 1.1 $\mu$m, beyond the wavelength range shown.}
\label{fig:geometric_albedo}
\end{center}
\end{figure}

\subsection{Geometric albedo}
Measurements of the geometric albedo of HD~189733b \citep{EvansEtAl2013,KrennEtAl2023HD189CHEOPS}  could also provide constraints on the haze production rate and optical properties.  In Fig. 
\ref{fig:geometric_albedo}, we compare the geometric albedo predicted from our simulations to observations. At short wavelengths ($<0.5 \,\mu$m), both soots and Titan-type hazes efficiently absorb incoming starlight, leading to a stark decrease in the albedo compared to the clear-atmosphere model. Soots are more absorbing than Titan-type hazes, resulting in a geometric albedo that is lower by a factor of about two for the same haze production rate ($2.5\cdot10^{-11}$ kg m$^{-2}$ s$^{-1}$) for these short wavelengths. For wavelengths $>0.5 \,\mu$m, sodium absorption dominates. Thus, the geometric albedo is very low for all simulations, including the clear and all hazy cases.

Overall, observational constraints from albedo spectra prefer a clear atmosphere or low haze production rates. The clear atmosphere model as well as the soot model with the lowest haze production rate match the observations with the HST STIS G430L grating \citep{EvansEtAl2013} reasonably well, while all simulations with higher haze production rates (soots and Titan-type hazes) result in a too low albedo at short wavelengths and a less pronounced drop from short wavelengths to longer wavelengths. All simulations underpredict the albedo in the CHEOPS bandpass, though again, the clear-atmosphere- and low-haze-production models are closer to the observations.

\subsection{Emission spectra}
\begin{figure*}
\begin{center}
\plotone{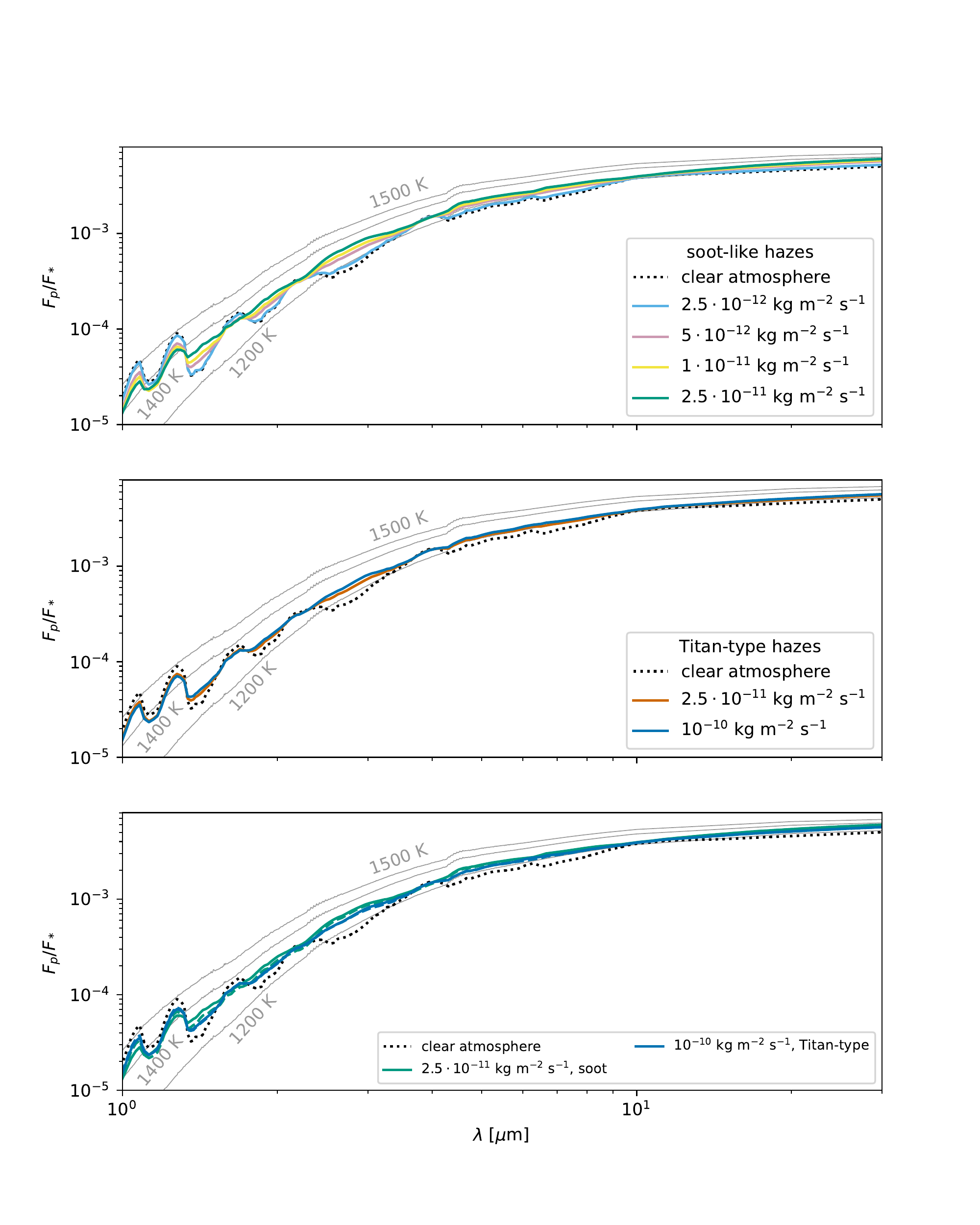}
\caption{Model-predicted emission spectra. For comparison, blackbody emission spectra at several temperatures are shown as thin gray lines. In the bottom panel, dashed lines show spectra in which the haze opacity was neglected during post-processing, thus isolating the effect of the changed thermal structure.}
\label{fig:emissionspectra_models}
\end{center}
\end{figure*}

\begin{figure*}
\begin{center}
\plotone{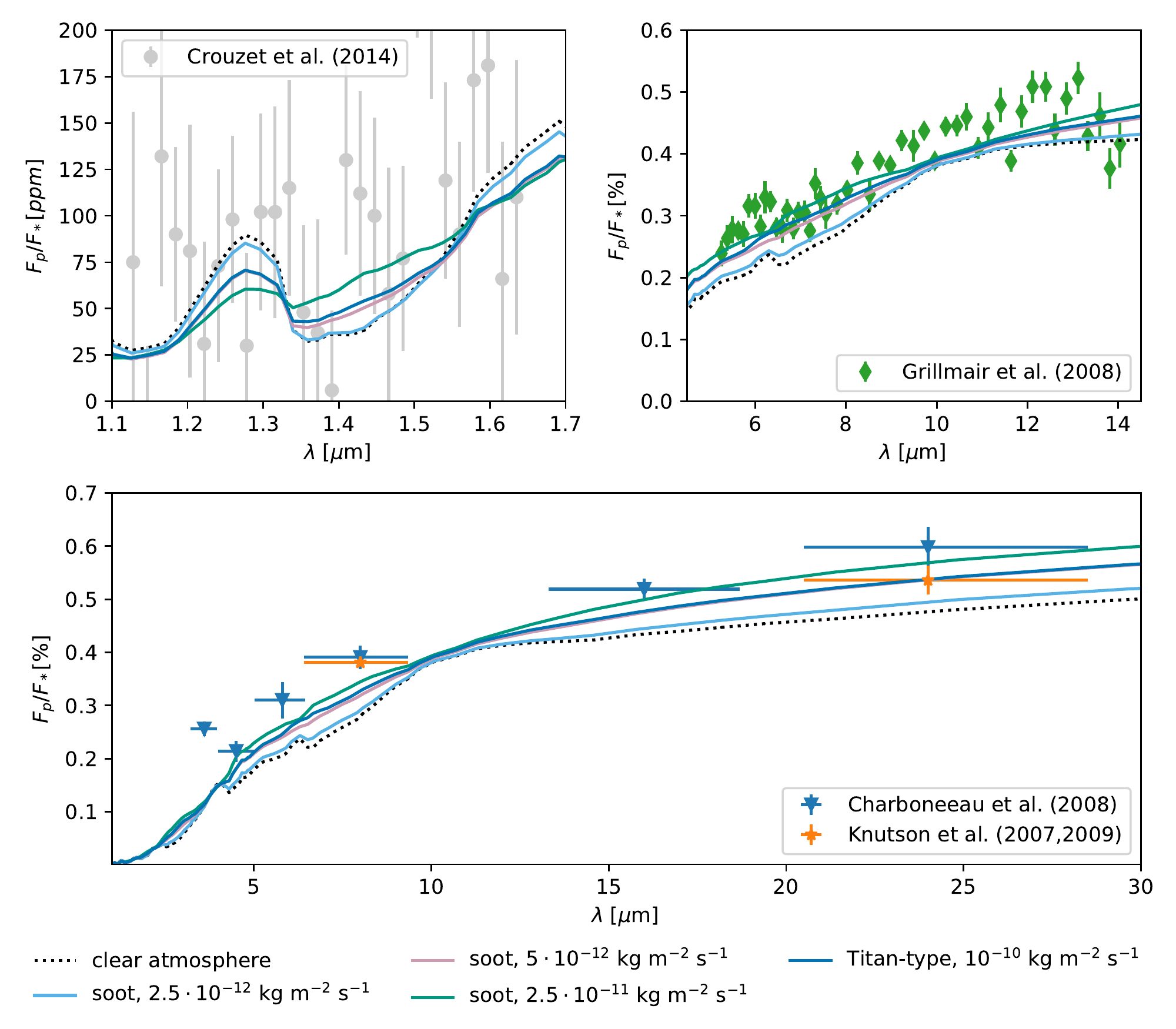}
\caption{Comparison of selected model-predicted emission spectra to a range of observations. Upper left: HST WFC3 G141 grism observations \citep{CrouzetEtAl2014}. The error bars include the combined uncertainties of the differential spectrum and the white light curve. Upper right: Spitzer IRS observations \citep{GrillmairEtAl2008}. Bottom: Spitzer IRAC \citep{CharbonneauEtAl2008HD189733b,KnutsonEtAl2007HD189phasecurve}, IRS photometry \citep{CharbonneauEtAl2008HD189733b} and MIPS \citep{CharbonneauEtAl2008HD189733b,KnutsonEtAl2007HD189phasecurve}  observations.
}
\label{fig:emissionspectra_observations}
\end{center}
\end{figure*}

Haze radiative feedback leads to changes in the emission spectrum, in most cases reducing the amplitude of spectral features in the near-infrared water bands and increasing the flux at long wavelengths ($>$ 4 $\mu$m) (Fig. \ref{fig:emissionspectra_models}). These changes are mostly driven by the changes in the temperature structure of the atmosphere due to haze radiative feedback rather than the addition of haze opacity when calculating the transmission spectrum (bottom panel). Between 1 and 2 $\mu$m, the emission is probing relatively deep layers of the atmosphere (up to 1~bar outside the water bands, $\approx$50-100~mbar inside the water bands), below the thermal inversions. Water is therefore seen in absorption. In this pressure region, simulations with haze radiative feedback exhibit a much smaller temperature gradient than the clear atmosphere simulation, leading to a reduced amplitude of the water feature. Especially in the soot-like simulation with the highest haze production rates ($2.5\cdot10^{-11}$ kg m$^{-2}$ s$^{-1}$), it appears that the emission is probing the region in which the temperature profile transitions from decreasing with altitude to a thermal inversion. In this transition region, the temperature profile is close to isothermal, leading to a particularly low feature amplitude. Between 2 and 3 $\mu$m, the soot-like simulations with intermediate haze production rates and both Titan-type simulations show a spectrum close to that of a blackbody. The soot-like simulation with the highest haze production rate exhibits an emission feature, while the same feature is seen in absorption in the simulation with the lowest haze production rate. At wavelengths beyond 4 $\mu$m, all models with haze feedback emit more radiation than the haze-free simulation.

Comparing to existing observations (Fig. \ref{fig:emissionspectra_observations}) remains somewhat inconclusive. The \textit{HST WFC3} data from \citet{CrouzetEtAl2014} cannot distinguish between models with or without thermal inversion. All models are consistent with this observation. None of the models match the longer-wavelength observations well, neither the haze-free one nor the ones with haze feedback. In particular, the IRAC 3.6, 5.8 and 8 $\mu$m points \citep{KnutsonEtAl2007HD189phasecurve, CharbonneauEtAl2008HD189733b} deviate from the models by much more than their one-sigma error bars. All Spitzer observations show more emitted flux at secondary eclipse than our model spectra. Including haze radiative feedback increases emission at these wavelengths and thus moves the models somewhat closer to the observed flux.	However, at the same time, the water absorption feature between 6 and 8 micron observed with \textit{Spitzer IRS} \citep{GrillmairEtAl2008} disappears in most models with haze radiative feedback, with exception of the model with the weakest haze production rate. With the current quality of observations, the latter is indistinguishable from the haze-free model in emission despite exhibiting two substantial thermal inversions.

We note that the models presented in this study are pure forward models and we did not actively attempt to improve the match to secondary eclipse observations. To improve the match, exploring additional parameters such as metallicity and atmospheric drag would be necessary.

\subsection{Phase Curves}
\begin{figure*}
\begin{center}
\includegraphics[width=0.8\textwidth]{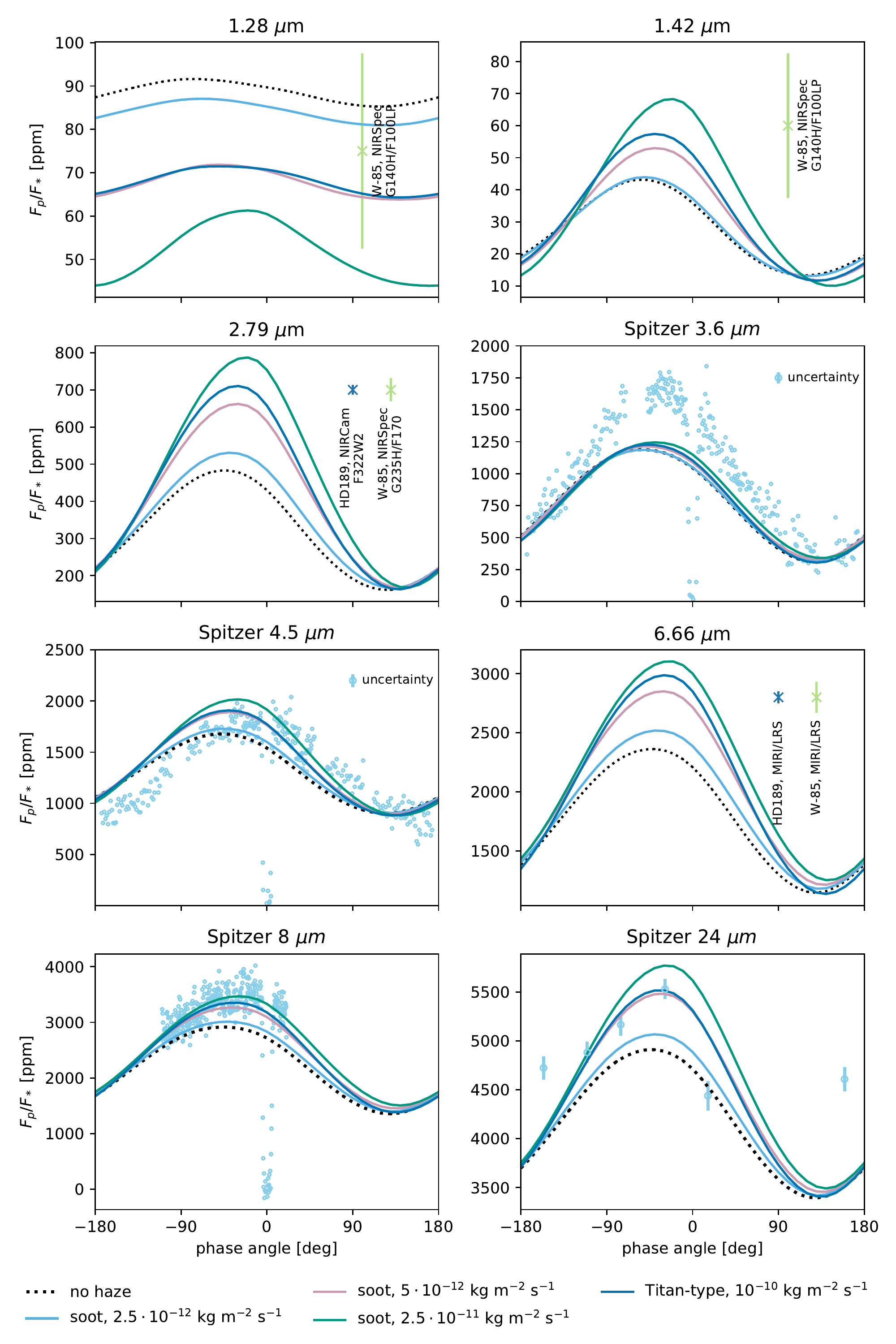}
\caption{Model-predicted phase curves at different wavelengths and in multiple Spitzer bandpasses. The observational data in the Spitzer bandpasses (light blue circles) are the same as in Fig. 12 in \cite{KnutsonEtAl2012} and are taken from \cite{KnutsonEtAl2012} (3.6 and 4.5 $\mu$m), \cite{KnutsonEtAl2007HD189phasecurve,KnutsonEtAl2009_HD189,AgolEtAl2010} (8 $\mu$m) and \cite{KnutsonEtAl2009_HD189} (24 $\mu$m). In addition, estimated JWST errorbars are shown for both HD 189733b and WASP-85Ab, a planet with comparable properties orbiting a fainter star (see text). The errorbars were calculated for bin widths of 720~s, chosen to be identical to the bins in the Spitzer 3.6~$\mu$m and 4.5~$\mu$m observations panels.
}
\label{fig:phasecurves}
\end{center}
\end{figure*}
As demonstrated in Fig. \ref{fig:phasecurves}, haze radiative feedback can substantially affect thermal phase curves over a broad range of wavelengths. At most infrared wavelengths, including all wavelengths larger than 4.2 $\mu$m, the dayside flux increases drastically, while the nightside flux remains relatively unchanged. This leads to an overall increase in the phase curve amplitude. At the same time, the eastward offset of the phase curve decreases due to the changes in the temperature structure.
In the near-infrared, however, there are also multiple wavelength regions, in which the planetary flux decreases in the simulations with haze feedback at all phases (0.79-1.32 $\mu$m, 1.51-1.74 $\mu$m, 2.03-2.28 $\mu$m, 3.7-4.1 $\mu$m). These wavelength regions are atmospheric windows, in which the emission emerges from particularly deep regions. Thus, the emission is probing the regions on which hazes have as cooling effect.
Notably, in the Spitzer 3.6 $\mu$m band, haze feedback has little effect on the phase curve. The reason is that for some wavelengths within the bandpass (3.1-3.5 $\mu$m), haze feedback leads to more emission, while for other wavelengths (3.7-4.0 $\mu$m), haze feedback decreases the amount of emitted flux. Near the center of the bandpass, there is little difference between the phase curves. As a result, the band-averaged phase curves look similar for all simulations.

To motivate future phase curve observations with JWST, we calculated JWST errorbars using the open-source software \texttt{PandExo} \citep{BatalhaEtAl2017}  for the generated phase curves of HD 189733\,b  for the wavelengths shown in Fig. \ref{fig:phasecurves} (1.28, 1.42 , 2.79 and 6.66 $\mu$m). Due to the relatively high brightness of the host star (J = 6.07 mag), the NIRSpec Grisms, NIRSpec PRISM and NIRISS/SOSS oversaturate. Only the NIRCam Grisms and MIRI/LRS stay below the statuation limit and can be used to perform observations at the redder wavelengths. The only available spectroscopic modes to observe the planet at 2.79 and 6.66 micron are the NIRCam F322W2 and MIRI/LRS instruments, respectively. The resulting errorbars for these modes are included in Figure \ref{fig:phasecurves}. 

For the two shorter wavelengths, we simulated JWST observations for a planet similar to HD 189733 b with a fainter host star to prevent oversaturation of the detectors. We searched the NASA Exoplanet Archive\footnote{\url{https://exoplanetarchive.ipac.caltech.edu/}} for a planet with a similar orbital period, equilibrium temperature, mass and radius to HD 189733 b orbiting a fainter host star. The hot jupiter WASP-85A b (J = 9.28 mag) compares well with HD 189733 b and we summarize some fundamental parameters between the two systems in Table \ref{tab:Wasp85Ab}. Figure \ref{fig:phasecurves} shows the simulated JWST errorbars of WASP-85A b in all four wavelengths of interest. The highest precision for the 1.28 and 1.42 micron wavelengths is reached with NIRSpec G140H/F100LP. NIRSpec G235H/F170 provides the highest precision for the 2.79 micron bin. The 6.66 micron phase curve can only be observed with MIRI/LRS. 

\begin{deluxetable}{lrr}
\tablecaption{Comparison between HD 189733 b and WASP-85A b}
\label{tab:Wasp85Ab}
\tablehead{\colhead{Parameter} & \colhead{Value\tablenotemark{1}} & \colhead{Value\tablenotemark{2}} \\ 
\colhead{} & \colhead{HD 189733 b} & \colhead{WASP-85A b} } 
\startdata
		Period (days)               & 2.2186 & 2.6557 \\ %
		Radius ($R_{\textrm{Jup}}$) & 1.119	 & 1.24  \\ %
		Mass ($M_{\textrm{Jup}}$)       & 1.163  & 1.265  \\%
		Equilibrium temperature (K) & 1209   & 1443 \\%
		J (mag)                     & 6.07   & 9.28 \\
		K (mag)                     & 5.54   & 8.73 \\
		TSM\tablenotemark{3}        & 843    & 177 \\
		ESM\tablenotemark{3}        & 1117   & 224 \\
\enddata
\tablenotetext{1}{\citet{Addison2019}}
\tablenotetext{2}{\citet{Mocnik2016}}
\tablenotetext{3}{Transmission spectroscopy metric (TSM) and emission spectroscopy metric (ESM) following \citet{Kempton2018}.}
\end{deluxetable}

\section{Discussion}
\label{sec:discussion}
\subsection{Compatibility between models and different types of observations}
After comparing our simulations to various observations of HD 189733b in the previous section, it is obvious that there is no simulation that explains all of the observations well. The low-resolution emission data, in particular the IRS data, support a non-inverted temperature profile in the pressure regions probed in these observations, most consistent with a low haze production rate or clear atmosphere. The geometric albedo spectrum similarly precludes large haze production rates with soot-like or Titan-type hazes, as either are too absorbing at short wavelengths. A clear atmosphere or low haze production rate still results in a somewhat lower geometric albedo than observations but provides a much better match than large haze production rates.

However, the short-wavelength slope requires aerosols \citep[even when contamination from star spots is considered, e.g.,][]{ZhangEtAl2020HD189Retrieval,ArfauxLavvas2022}. Microphysics models of condensate clouds tend to form large particle sizes and thus struggle to reproduce the short-wavelength slope \citep{PowellEtAl2018,LinesEtAl2018TransmissionSpectra}. Photochemical hazes thus remain the most likely candidate for explaining the short-wavelength slope. In our simulations, Titan-type hazes produce a better match to the slope than soots, but require an offset between the optical and NIR data. In addition, the \textit{WFC3} transmission spectrum requires a source of near-infrared opacity to mute the water feature, either hazes or condensate clouds. For a solar metallicity, the haze production rates necessary to explain the low amplitude of the water feature are in conflict with the constraints from the geometric albedo. While we did not simulate super-solar metallicities, we expect that a higher metallicity would exacerbate this problem, because even more hazes would be needed to reduce the amplitude of the larger water feature.

Finally, observations using high-resolution cross-correlation have detected carbon monoxide \citep{deKokEtAl2013HD189CODayside,RodlerEtAl2013HD189CODayside} and water \citep{BirkbyEtAl2013HD189} in absorption in the dayside spectrum of HD 189733b, implying a temperature profile decreasing with height. This observation has seemingly caused tension with photochemical hazes producing a thermal inversion at high altitudes. The pressure probed in these observations cannot be directly constrained because the information of the continuum emission is lost in the process of removing telluric and stellar lines. However, using forward models, \citet{deKokEtAl2013HD189CODayside} estimate that the pressure probed by the CO lines is between $10^{-5}$ and $10^{-3}$ bar for a CO volume mixing ratio of $10^{-4}$. Our soot-like models naturally exhibit a decreasing temperature profile with height in this pressure region due to the low haze mass mixing ratio below the haze production region caused by upwelling on the day side. The high-resolution observations thus do not necessarily rule out a temperature inversion by soot-like hazes.

The remaining tension between different types of observations thus is that low-resolution secondary eclipse measurements (reflected light and thermal emission) support a low haze production rate, while the observed transmission spectrum requires models with substantial haze opacity in the near-infrared water bands and thus higher haze production rates. A potential way to reconcile this tensions could be models combining photochemical hazes to explain the short-wavelength slope with larger condensate clouds to explain the muted amplitude of the near-infrared water feature \citep[as already pointed out in][]{PontEtAl2013}. The potential impact of such condensate clouds on the simulation results is discussed further below. Another possibility is that the photochemical hazes in HD 189733b’s atmosphere could have a lower absorption cross-section in the UV than the soot and Titan-type hazes we considered. Such hazes, possibly in combination with reflecting condensate clouds deeper in the atmosphere, could increase the geometric albedo and thus bring models into better agreement with the optical secondary eclipse measurements. Recently, new refractive indices derived from laboratory experiments simulating haze formation in conditions relevant to super-Earths showed substantially less UV absorption than Titan-type hazes \citep{HeEtAl2023WaterRichOpticalProperties,CorralesGavilanEtAl2023}. However, less absorption in the UV may come at the cost of a worse match to the short-wavelength slope seen in transmission.

\subsection{Importance of wavelength-dependent radiative transfer}
The comparison of double-gray and correlated-k radiative transfer in Section \ref{sec:grayvssparc} highlights the importance of the choice of radiative transfer.
For the purpose of studying photochemical hazes, it appears necessary to use the more computationally expensive correlated-k radiative transfer. This represents a major challenge for future larger parameter studies. It may be worth evaluating how well radiative transfer schemes with a complexity level between double-gray and correlated-k, for example the picket-fence scheme in \citet{LeeEtAl2021PicketFenceGCM}, can reproduce the haze distribution from the correlated-k approach.

\subsection{Choice of haze optical properties}
Our results further demonstrate that the assumed optical properties of hazes strongly influence atmospheric dynamics. The strength and shape of the equatorial jet strongly differs between the two different assumed haze refractive indices. The resulting 3D distribution of the haze mass mixing ratio also looks dramatically different. Hazes with a soot-like refractive index are more concentrated at the nightside and morning terminator than at the evening terminator, while hazes with a refractive index similar to Titan-type hazes are more concentrated at the evening terminator.

Currently, there are little experimental and theoretical constraints on the optical properties of photochemical hazes in hot Jupiter atmospheres. Measured refractive indices are either derived from soots produced in hydrocarbon flames or from experiments simulating haze formation on Titan, conducted in a nitrogen-dominated atmospheres either at room temperature or at Titan-like temperatures ($\approx 100$~K).
In recent years, several research groups have produced haze analogs intended to simulate conditions on exoplanets \citep{HorstEtAl2018,FleuryEtAl2019,GavilanEtAl2018}.
Out of these experiments, refractive indices have only been published for temperate water-dominated \citep{HeEtAl2023WaterRichOpticalProperties} and nitrogen-dominated \citep{CorralesGavilanEtAl2023} atmospheres. Both sets of refractive indices show substantial deviations from the \citet{KhareEtAl1984} tholins. The differences include the absolute value of the complex refractive index,  additional spectral features due to incorporated oxygen, and location of the ``spectral window'' with a low imaginary refractive index in the optical-to-near-IR. 

Refractive indices for hydrogen-dominated, high-temperature haze analogs have not been published so far. However, the color of haze analogs produced in the experiments by \citet{HorstEtAl2018} strongly depends on the temperature and gas composition of the atmosphere \citep{HeEtAl2018}. In addition, these haze analogs have incorporated more oxygen than Titan haze analogs or soots \citep{MoranEtAl2020}. Note that their initial experiments only cover temperatures up 600~K, with a more recent update including 800~K\citep{HeEtAl2020}, not hot enough to match the temperatures in the haze production region of HD~189733b. \citet{FleuryEtAl2019} report the formation of solid photochemical products in an experiment at 1,473~K in a hydrogen-dominated gas mixture with a C/O ratio of 1. Their haze analogs show infrared spectral signatures of carbonyl and aldehyde groups, indicating a solid composition based on carbon, oxygen and hydrogen compounds. However, while the detection of carbonyl and aldehyde spectral signatures is evidence of the incorporation of oxygen into the high-temperature haze analogs, no statement about the relative oxygen content can be made from these measurements alone.

The optical properties of photochemical hazes in the atmospheres of hot Jupiters and hot Neptunes thus could substantially deviate from both the soot-like and the Titan-type refractive indices used in our simulations. It is also possible that these optical properties vary substantially between individual planets with different temperatures, around different stellar types or with varying atmospheric composition. Given the dramatic effect of the haze refractive index on atmospheric dynamics and the 3D distribution of hazes, we stress the need for measurements of the refractive indices of laboratory haze analogs specific to hot hydrogen-dominated atmospheres.

\subsection{Changes to chemistry and haze production rate due to haze feedback}
We note that all the presented simulations have a fixed haze production rate. In a real atmosphere, however, the changed temperature structure and circulation will affect chemical processes in the atmosphere and thus the haze production rate. This effect is non-local and therefore difficult to model: On one hand, the hotter temperatures at low pressures directly affect the photochemical reactions producing haze precursor molecules. On the other hand, the cooler temperatures in the region where methane is quenched between 100 mbar and 10 bar change the amount of methane that is mixed upwards to the photochemically active regions. This will affect the abundance of all methane-derived photochemical products and thus many haze precursor species. \citet{ArfauxLavvas2022} found in 1D simulations that the changed temperature structure due to haze radiative feedback can alter haze production rates by a factor of a few.
If the hazes are as refractory as soot, evaporation or thermal decomposition of the hazes at low pressures due to the haze-induced thermal inversion is not anticipated, as the temperature would have to exceed 1,800~K \citep[see e.g., Fig.~2 in ][]{LavvasKoskinen2017}.

\subsection{Impact of condensate clouds}
Finally, it is likely that in the atmospheres of many hot Jupiters, including HD 189733b, both photochemical hazes and condensate clouds are present. Condensate clouds are likely to further alter temperature structure and atmospheric circulation \citep[e.g.,][]{LeeEtAl2016,LinesEtAl2018,RomanRauscher2019}. 
They can also substantially influence infrared phase curves. Notably, partial cloud coverage is known to decrease the phase curve offset at infrared wavelengths by blocking emission from deeper layers of the atmosphere near the evening terminator \citep{ParmentierEtAl2021,RomanEtAl2021}. Photochemical hazes do not produce this effect due to their much smaller particle sizes, which make them more transparent in the infrared. Including condensate clouds therefore could further improve the agreement with phase curve observations. We also note that in our simulations, photochemical hazes did not substantially lower the night side fluxes. Condensate clouds are therefore still the favored explanation for the observed uniformly low nightside temperatures of hot Jupiters \citep{BeattyEtAl2019KELT-1bNightsideClouds,KeatingCowan2019}. Many real atmospheres likely include both types of aerosols and future models including both are encouraged.

It is also likely that condensate clouds interact with photochemical hazes, for example by condensing onto photochemical haze particles. While we consider the removal of haze particles from the distribution of ``purely photochemical'' hazes in a highly idealized fashion (through the sink term), detailed microphysical and laboratory studies of the interactions between photochemical hazes and condensate clouds, similar to \citet{YuEtAl2021HazeSurfaceEnergy}, in conditions expected in hot Jupiter atmospheres are desirable.

\section{Conclusion}
\label{sec:conclusion}
In this work, we examined the effect of radiative feedback of photochemical hazes on temperature structure, atmospheric circulation and haze distribution in a 3D general circulation model (GCM) of hot Jupiter HD~189733b using a state-of-the-art GCM with wavelength-dependent (correlated-k) radiative tranfer. First, we performed a detailed comparison of temperature structure, circulation and the distribution of radiatively passive hazes between double-gray and correlated-k radiative transfer.
Compared to the double-gray model, the correlated-k simulation has lower temperatures and a stronger day-night temperature contrast at low pressures. There are further changes to the structure of the equatorial jet, which is narrower and broadens less with height in the correlated-k simulation, the location of the mid-latitude nightside vortices and the regions of strongest downwelling. These changes lead to the haze mass mixing ratio peaking along a ring surrounding the center of the nightside vortices in the correlated-k simulation rather than in the center of the vortices.
The mass mixing ratio also drops off faster with increasing pressure in the correlated-k simulation.

Then, we performed simulations with correlated-k radiative transfer that included heating and cooling by photochemical hazes. Hazes in our model are an active tracer that dynamically interacts with atmospheric circulation. The majority of simulations assumed a soot-like complex refractive index, but we also explored a refractive index resembling Titan-type hazes.
In both cases, a strong temperature inversion forms at low pressures on the dayside. Therefore, the day-to-night temperature contrast increases dramatically for p$<$10~mbar (400--700~K instead of 200~K in the passive-haze simulation). The detailed structure of the dayside temperature profile differs between soot-like and Titan-type hazes: For soot-like hazes, there are two separate temperature maxima (near 1 µbar and 1 mbar, respectively), separated by a temperature minimum at 10 µbar (just below the haze production region). The temperature minimum is caused by upwelling on the dayside mixing haze-poor air upwards, an effect not captured in 1D simulations. For Titan-type hazes, the dayside-average temperature profile is more uniform with a large isothermal region at low pressures.

The response of atmospheric circulation to heating and cooling from photochemical hazes strongly depends on the choice of the complex refractive index of the haze particles. For soot-like hazes, which are highly absorptive and exhibit a weak wavelength-dependence of the absorption cross section, the equatorial jet slows down and broadens at low pressures. Vertical velocities increase, especially upwelling near the terminator and on the dayside. The higher the haze production rate, the stronger are these changes. There are only moderate changes to the haze distribution compared to radiatively passive hazes.

For hazes with a refractive index similar to Titan-type hazes, the equatorial jet accelerates substantially, especially at low pressures. This results in eastward velocities throughout the entire dayside in the haze production region (in contrast to polewards and eastward velocities in the radiatively passive and soot-like simulations). Hazes thus are effectively homogenized across most of the globe with the exception of the nightside mid-latitude vortices, which remain depleted of hazes. The distribution of hazes thus strongly contrasts from that of radiatively passive or soot-like hazes, with overall more hazes at the evening terminator than at the morning terminator.

We suggest that the difference between soot-like and Titan-type-haze simulations is caused by the different vertical heating profiles. For Titan-type hazes, the stellar heating is more spread out vertically, leading to a stronger net heating at higher pressures, that then drives a stronger equatorial jet.
Due to this unexpectedly strong dependence of the 3D haze distribution and atmospheric circulation on the assumed haze optical properties, we emphasize the need for better constraints on the refractive indices of photochemical hazes under conditions relevant to hot Jupiters, for example by measurements of the refractive indices of laboratory haze analogs formed at high temperatures in hydrogen-dominated atmospheres. 

Including haze radiative feedback does not improve the match to transmission spectra. Alternative explanations for the steep short-wavelength slope, such as star spots, an additional offset between the STIS and WFC3 data \citep[as discussed in][]{ArfauxLavvas2022}, different haze optical properties, or strong sub-grid-scale mixing \citep{SteinrueckEtAl2021} are required to explain observations.

In emission, haze radiative feedback leads to a decreased amplitude of the near-infrared water features and increased emission at wavelengths past $\approx$ 4 $\mu$m. In most wavelength regions, the phase curve amplitude increases substantially, while the eastward phase curve offset is reduced. Notably, the soot-like simulation with the lowest haze production rate is almost indistinguishable from the haze-free model in emission despite exhibiting a moderate double-peaked thermal inversion.

We point out that current models (neither 1D nor 3D, whether clear-atmosphere, photochemical hazes or condensate clouds) still do not fully explain the set of existing observations of this benchmark hot Jupiter. While Titan-type hazes provide a better (though imperfect) match to transit observations, the detection of CO in absorption in high-resolution observations favors soot-like hazes because the double-peaked temperature structure in soot-like simulations is compatible with a declining temperature profile with height in the pressure range likely probed in these observations. In addition, geometric albedo constraints from optical secondary eclipse measurements prefer low haze production rates that would require an additional opacity source such as condensate clouds to explain the low amplitude of the near-infrared water band in transmission.

\begin{acknowledgments}
M.S. was supported by NASA Headquarters under the NASA Earth and Space Science Fellowship Program - Grant 80NSSC18K1248 while working as a graduate student at the University of Arizona. X.Z. acknowledges support from NASA Interdisciplinary Consortia for Astrobiology Research (ICAR) grant 80NSSC21K0597 and the NASA Exoplanet Research Grant 80NSSC22K0236. We thank Thaddeus Komacek for conversations on the bottom boundary condition and Peter Gao for sharing the model grid used in \citet{ThorngrenEtAl2019InternalTemperature}. 
M.S. further thanks Elspeth Lee for discussions on the comparison between double-gray and correlated-k radiative transfer, Maria Zamyatina and Kazumasa Ohno for discussions on the haze optical properties, and Gilda Ballester for helpful comments on an early draft. We also thank the referee for helpful suggestions that improved the manuscript substantially. This research made use of NASA's Astrophysics Data System. This research has made use of the NASA Exoplanet Archive, which is operated by the California Institute of Technology, under contract with the National Aeronautics and Space Administration under the Exoplanet Exploration Program.
\end{acknowledgments}

%

\vspace{5mm}


\software{Numpy \citep{NumpyCitation},
          SciPy \citep{SciPyCitation}, 
          Matplotlib \citep{MatplotlibCitation},
          Cartopy \citep{Cartopy},
          Bibmanager \citep{Cubillos2020zndoBibmanager},
          Pandexo \citep{BatalhaEtAl2017},
          Astropy \citep{Astropy2022}
          }




\bibliography{biblio}{}
\bibliographystyle{aasjournal}


\end{document}